\documentclass[]{aa}  

\usepackage{graphicx}
\usepackage{txfonts}
\usepackage{booktabs}

\usepackage[]{hyperref}
\begin{document}

    \title{Evolution of solar wind sources and coronal rotation driven by the cyclic variation of the Sun's large-scale magnetic field}

   \titlerunning{Cycle to cycle variation}
   \authorrunning{Finley \& Brun}

   \author{A. J. Finley\inst{1}
          \and
          A. S. Brun\inst{1}
          }

    \institute{Department of Astrophysics-AIM, University of Paris-Saclay and University of Paris, CEA, CNRS, Gif-sur-Yvette Cedex 91191, France \\ \email{adam.finley@cea.fr} }

   \date{Received July 30, 2023; accepted Sept 19, 2023}


\abstract{The strength and morphology of the Sun's magnetic field evolves significantly during the solar cycle, with the overall polarity of the Sun's magnetic field reversing during the maximum of solar activity. Long-term changes are also observed in sunspot and geomagnetic records, however systematic magnetic field observations are limited to the last four cycles.}
{Here, we investigate the long-term evolution of the Sun's magnetic field, and the influence this has on the topology and rotation of the solar corona.}
{The Sun's photospheric magnetic field was decomposed into spherical harmonics using synoptic Carrington magnetograms from 1) the Wilcox Solar Observatory, 2) the Michelson Doppler Imager onboard the Solar and Heliospheric Observatory, and 3) the Helioseismic and Magnetic Imager onboard the Solar Dynamics Observatory. The time-evolution of the spherical harmonic coefficients was used to explore the variation of the Sun's magnetic field, focusing on the large-scale modes. Potential field source surface extrapolations of the photospheric field were computed to follow topological changes in the corona.}
{The footpoints of the Sun's open magnetic field vary between the polar coronal holes and activity driven features such as active regions, and equatorial coronal holes. Consequently, the mean rotation rate of the solar wind is modulated during each cycle by the latitudinal variation of open field footpoints, with slower rotation during minima and faster (Carrington-like) rotation during maxima.}
{This variation is sensitive to cycle to cycle differences in the polar field strengths and hemispherical flux emergence rates, with the ratio of quadrupole to dipole energy following a similar variation. Cycle 23 maintained a larger fraction of quadrupolar energy in the declining phase, which kept the sources of open magnetic flux closer to the equator, extending the period of faster equator-ward connectivity. The ratio of quadrupole to dipole energy could be a useful proxy when examining the impact of differential rotation on the coronae of other Sun-like stars.}


   \maketitle
%

\section{Introduction}

The Sun's magnetic field undergoes an approximately 11 year cycle of activity \citep[see review of][]{hathaway2015solar}. Beginning with a mostly axisymmetric dipolar magnetic field at solar minima \citep{derosa2012solar}, the Sun's magnetic field becomes increasingly complex as the cycle progresses due to the emergence of active regions \citep[the properties of which are reviewed in][]{van2015evolution}. These regions are typically bipolar in nature, following Hale's law for leading-trailing polarity and Joy's law for tilt angle (summarised in \citealp{hale1919magnetic}, and more recently \citealp{dasi2010sunspot}). The presence of strong magnetic fields in these regions suppresses surface convection, creating dark patches on the solar disk, called `sunspots', which have been well-documented throughout the last four centuries \citep[see][]{clette2014revisiting}. Flux emergence typically begins at mid-latitudes, steadily moving down towards the Sun's equator as activity increases \citep{carrington1858distribution, sporer1880beobachtungen, hathaway2003evidence}. The emerging field interacts with pre-existing structures in the low-corona, causing magnetic energy to build-up, then be released in the form of flares \citep{toriumi2019flare}, and coronal mass ejections \citep{forbes2000review}. The Sun's magnetic field reverts back to an axisymmetric dipole at the end of the activity cycle, however the magnetic polarity is reversed \citep{mordvinov2019evolution}. After two activity cycles ($\sim$22 years), the original magnetic field polarity is restored. Long-term trends in solar activity (on timescales of hundreds to thousands of years) are also observed when studying cosmogenic radionuclides stored in natural archives \citep[e.g.][]{usoskin2021solar}.

Features embedded in the photosphere are observed to differentially rotate \citep[see review of][]{beck2000comparison}, with the Sun's equator rotating faster than the poles. Typically one full revolution at the equator takes 24.5 days, in contrast to 33.4 days near the poles. Helioseismic inversions of the solar interior confirm that this differential rotation pattern permeates the entire convective zone, down to around $0.7 R_{\odot}$ \citep{thompson1996differential, schou1998helioseismic, larson2018global}. Below which there is a transition to rigid-body rotation in the radiative zone \citep{howe2009solar}. Active regions are broken apart by differential rotation over time \citep{gigolashvili2013investigation, imada2018effect}, with the exception of some very strong magnetic field regions which maintain a degree of cohesion \citep{yan2018successive}. Typically, active regions rotate near the Carrington rotation rate of 25.4 days (27.38 days as viewed from Earth), corresponding to the mean rotation rate at active solar latitudes.


The evolution of the Sun's magnetic field during each solar cycle leads to a cyclic variation in solar wind sources \citep{wang2009coronal}. This was explored in a range of previous theoretical works, using Potential Field Source Surface (PFSS) models \citep{stansby2021active}, magnetofrictional models \citep{yeates2014coronal}, and full magnetohydrodynamic simulations \citep{reville2017global}. During solar minima, the coronal magnetic field is mostly dipolar, with the solar wind emerging along open field at the rotational poles \citep[see also remote-sensing and in-situ observations][]{wilhelm1998solar, harvey2002polar, mccomas2008weaker}. As the Sun becomes more active, emerging active regions increase the complexity of the coronal field \citep{van2012magnetic}, allowing the solar wind to emerge from a broader range of sources, from around the active regions themselves \citep[recently investigated with Solar Orbiter by][]{yardley2023slow}, to equatorial coronal holes, and ephemeral regions. \citet{finley2023accounting} proposed that the evolving distribution of source regions drives a variation in the mean rotation rate of the corona during the solar cycle.

In this study, the methodology of \citet{finley2023accounting} is applied to a larger range of observations, spanning more than four solar cycles. Synoptic Carrington magnetograms were used to drive PFSS modelling, with each magnetogram consisting  of data assimilated during one Carrington rotation (CR) of the Sun (as viewed from Earth). The data products used in this work are summarised in Section 2. The decomposition of the magnetograms into a series of spherical harmonics is detailed in Section 3, from which trends in mode strengths were extracted (continued in Appendix \ref{ap2}). Section 4 presents the resulting latitudinal variation of solar wind sources from the PFSS modelling along with the impact on coronal rotation. The cycle to cycle variation of these quantities is discussed in Section 5.

\section{Observations}
Magnetograms were taken from the Wilcox Solar Observatory (WSO) and both the Michelson Doppler Imager (MDI), onboard the Solar and Heliospheric Observatory (SOHO), and the Helioseismic and Magnetic Imager (HMI), onboard the Solar Dynamics Observatory (SDO). Each individual Carrington magnetogram combines multiple full-disk line-of-sight magnetic field observations spanning a full CR of 27.38 days. The radial magnetic field component was derived from each line-of-sight measurements before assimilation, and the polar field strengths are corrected to account for poor visibility and high inclination.

It has been well-established that the magnetic field measurements from different instruments tend to disagree with one another \citep[e.g.][]{riley2014multi}. This often includes scale-dependent discrepancies that affect the energy recovered in each spherical harmonic mode \citep[as shown in][]{virtanen2017photospheric}. This is further complicated when comparing ground-based to space-based telescopes where the visibilities and spatial resolutions can vary significantly. For the period of overlap between WSO, SOHO/MDI and SDO/HMI (CR 2100 - 2107), a comparison of their relative field strengths was undertaken, detailed in Appendix \ref{ap1}. During this period, WSO field strengths were systematically smaller than their space-based counterparts\footnote{Previous works have shown that the WSO field strengths are around a factor 1.8 too low due to instrumental effects \citep{svalgaard1978strength}. This is included in the factor of 3.5 used to match the space-based observations.}, and SOHO/MDI field strengths were slightly larger than SDO/HMI \citep[see also the comparison in][]{liu2012comparison}. In order to produce a consistent timeseries, the WSO magnetograms were multiplied by a factor of $3.5$ and the SOHO/MDI magnetograms by a factor of $0.85$. These factors were included throughout our analysis, bringing the field strengths of the timeseries into agreement with magnetograms from SDO/HMI. This normalisation does not influence the spherical harmonic decomposition of the magnetograms, nor the connectivity of the PFSS models, used in this study.

\begin{figure}
 \centering
  \includegraphics[trim=0cm 0cm 0cm 0cm, clip, width=0.5\textwidth]{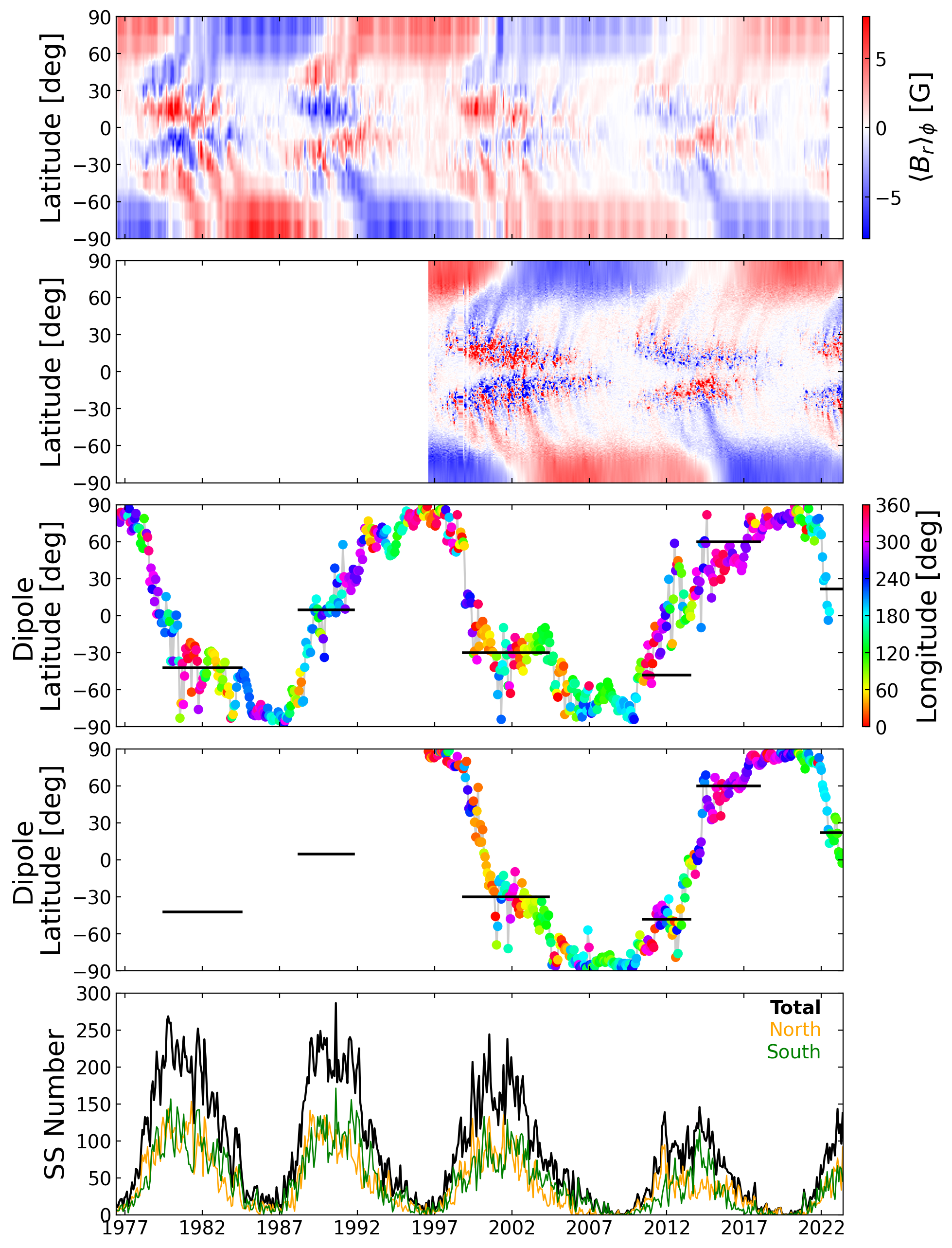}
   \caption{Evolution of the Sun's radial magnetic field from WSO and SOHO/MDI \& SDO/HMI timeseries. The top two panels show azimuthally averaged Carrington magnetograms from WSO (1976 - 2022) and from SOHO/MDI \& SDO/HMI (1996 - present). The following two panels show the latitude of the solar dipole (positive pole) deduced from the spherical harmonic-decomposition of the magnetograms from WSO, and SOHO/MDI \& SDO/HMI, respectively. The Carrington longitude of the dipole (positive pole) is then indicated in colour. Epochs where the dipole reversal appears to stall are highlighted with black horizontal bars. The bottom row displays the monthly sunspot number during this time, taken from WDC-SILSO, along with the contributions from the northern and southern hemispheres individually.}
   \label{fig:dipangle}
\end{figure}

\begin{table*}
\caption{Cycle lengths and lag times between maxima in each hemisphere.}
\label{table:cycles}
\centering
\begin{tabular}{c | c c c| c c c } 
\hline\hline
Cycle & Start & Sunspot Cycle & Magnetic Cycle & North Max. & South Max. & North-South Lag\\
 No. & [Decimal yr] & Length [yr] & Length [yr] & [yr after Min] & [yr after Min] & [yr]\\
\hline
12 & 1878.92 & 11.17 & -- & 2.75 & 5.0 & -2.25 \\
13 & 1890.08 & 11.83 & 23.0 & 2.41 & 3.5 & -1.08 \\
14 & 1901.92 & 11.58 & 22.75 & 4.08 & 5.33 & -1.25 \\
15 & 1913.5 & 9.67 & 20.84 & 4.08 & 4.08 & 0.0 \\
16 & 1923.16 & 10.5 & 21.67 & 5.34 & 4.84 & 0.5 \\
17 & 1933.67 & 10.5 & 21.67 & 3.83 & 4.83 & -1.0 \\
18 & 1944.16 & 10.08 & 21.25 & 5.67 & 3.08 & 2.59 \\
19 & 1954.25 & 10.34 & 21.51 & 5.0 & 3.42 & 1.58 \\
20 & 1964.58 & 11.58 & 22.75 & 4.58 & 5.5 & -0.92 \\
\hline
21 & 1976.16 & 10.5 & 21.67 & 3.5 & 4.08 & -0.58 \\
22 & 1986.67 & 9.75 & 20.92 & 3.0 & 4.83 & -1.83 \\
23 & 1996.42 & 12.5 & 23.67 & 4.08 & 5.67 & -1.59 \\
24 & 2008.92 & 11.01 & 22.18 & 2.83 & 5.17 & -2.34 \\
25 & 2019.92 & -- & -- & -- & -- & -- \\

\hline
\hline 
Range of Values & & 9.67 to 12.5 & 20.84 to 23.67 & 2.41 to 5.67 & 3.08 to 5.67 & -2.34 to 2.59 \\

Mean & & 10.85&21.99&3.94&4.56&-.62\\
\hline
\hline 
\end{tabular}
\end{table*}

\begin{figure}
 \centering
  \includegraphics[trim=0cm 0cm 0cm 0cm, clip, width=0.5\textwidth]{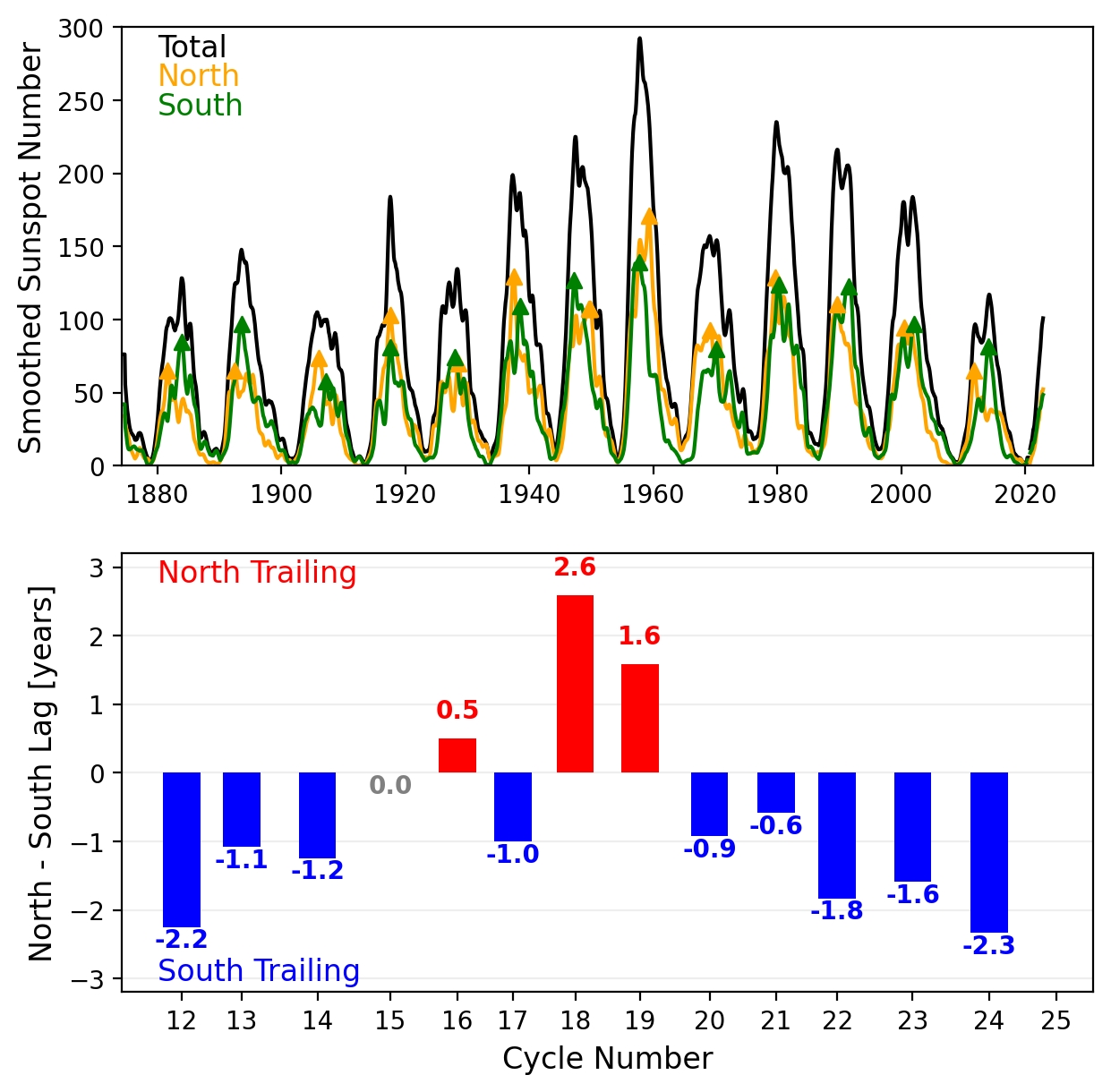}
   \caption{Smoothed 13-month total and hemispherical sunspot numbers. Maxima in each hemisphere are identified with triangular symbols. The lag times between hemispherical maxima are plotted in the lower panel for each cycle. Cycles 21-24, under investigation in this study, all had activity peaking in the northern hemisphere before the southern hemisphere.}
   \label{fig:cyclelags}
\end{figure}

\subsection{Wilcox Solar Observatory}
Magnetograms from the WSO \citep{scherrer1977mean}, used in this study\footnote{Data accessed Jan. 2023: http://wso.stanford.edu/synopticl.html}, span from 1976 to mid-2022 (CR 1641 - 2258). The WSO timeseries covers 46 years (617 CRs), over four sunspot cycles (21 - 24), and includes the rising phase of cycle 25. WSO magnetograms are available with a resolution in latitude and longitude of 36 and 72 points, respectively. Each data point therefore spans an area of $5^{\circ}$ x $5^{\circ}$ (60~Mm x 60~Mm). The evolution of the azimuthally averaged magnetic field from this timeseries is shown in the top panel of Figure \ref{fig:dipangle}.

\subsection{SOHO/MDI and SDO/HMI}
Magnetograms from SOHO/MDI \citep{scherrer1995solar} are available from 1996 to early 2011 (CR 1908 - 2107), with some data gaps, like the temporary loss of SOHO in 1998. Magnetograms from SDO/HMI \citep{scherrer2012helioseismic} are available from 2010 to present (CR 2100 - 2270). Both data products include a polar field correction \citep{sun2011new,sun2018polar}. For this study, SOHO/MDI magnetograms\footnote{Data accessed May 2023: http://hmi.stanford.edu/data/synoptic.html} are used from CR 1908 - 2099, and SDO/HMI, from CR 2100- 2270. The combined SOHO/MDI and SDO/HMI timeseries covers 27 years (361 CRs), just over half of the WSO timeseries, encompassing two sunspot cycles (23 and 24), and cycle 25 to present. SOHO/MDI \& SDO/HMI magnetograms are available with a resolution in latitude and longitude of 1800 and 3600 points, respectively. Each data point therefore spans an area of $1^{\circ}$ x $1^{\circ}$ (6~Mm x 6~Mm). The evolution of the azimuthally averaged magnetic field from this timeseries is shown in the second panel of Figure \ref{fig:dipangle}.

\subsection{Sunspot cycle lengths and hemispherical offsets}

Despite magnetogram records being limited to roughly the last four cycles, historical records of the sunspot number are available for the last four centuries \citep[e.g.][]{clette2014revisiting}. Of interest to this study are the hemispherical sunspot number records, here \citet{veronig2021hemispheric} is used. This record spans from May 1874 to present, from which cycle lengths and hemispherical asymmetry were extracted. Table \ref{table:cycles} contains the length of each sunspot cycle, magnetic cycle (defined as the current cycle length plus the previous one), timing of the maxima of activity in each hemisphere after solar minimum, and the relative lag time between each hemisphere's maximum. 

The mean sunspot cycle, and magnetic cycle, during this period are $10.85 \pm 0.82$ and $21.99 \pm 0.85$ years respectively, in-line with the standard quoted values, however individual cycles can have a significant deviation from this mean \citep[see also][]{wilson1987distribution, hathaway1994shape, hathaway2015solar}. For example, cycle 22 had a length of 9.8 years whereas cycle 23 was much longer with 12.5 years. This in turn allows for significant variation in consecutive magnetic cycles. (20.92 to 23.67). Hemispherical maxima were extracted from the monthly sunspot number records as smoothing with a 13-month window, shown in Figure \ref{fig:cyclelags} along with a depiction of the hemispherical lag times for each cycle. The maxima of the northern and southern hemispheres can have as much as 2.6 years of difference. Cycles 21-24, which are the focus of this study, have activity in the northern hemisphere systematically peaking before the southern hemisphere \cite[see also][]{deng2016systematic}. This is due to a significant quadrupolar mode, previously discussed in \citet{derosa2012solar}, see their appendix on dynamo symmetries. The current understanding is that the quadrupolar (symmetric) dynamo mode is playing a key role in off-setting the dipolar (antisymmetric) dynamo mode, leading to one hemisphere to be ahead or behind the other one depending on the relative signs of the two modes. The nonlinear coupling of the two dynamo modes make it difficult to predict which hemisphere will be ahead. The evolution of these two modes is discussed in Section \ref{quadrupole}.

\begin{figure}
 \centering
  \includegraphics[trim=0cm 0cm 0cm 0cm, clip, width=0.5\textwidth]{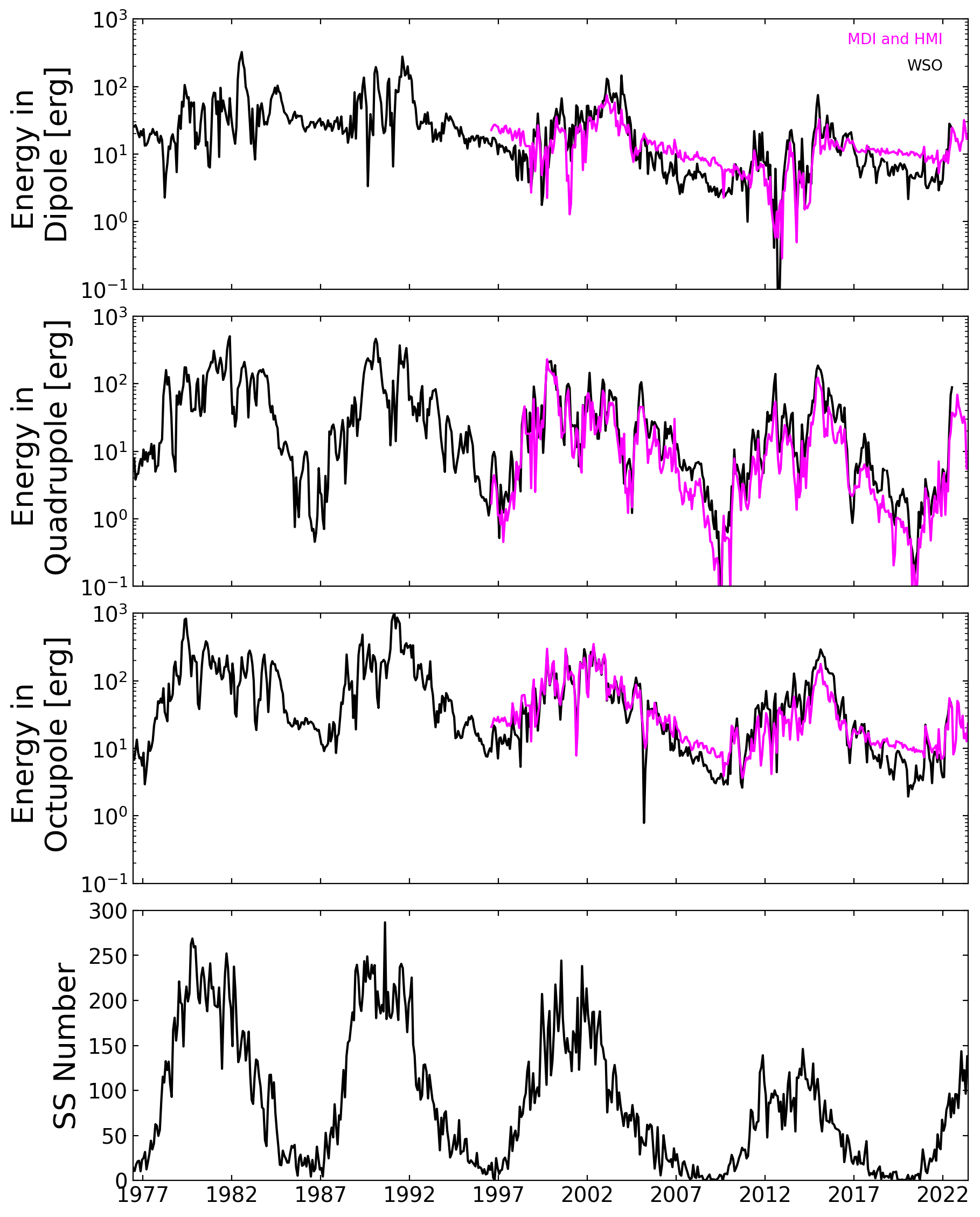}
   \caption{Comparison of energy in the dipole ($l=1$), quadrupole ($l=2$), and octupole ($l=3$) modes versus solar cycle. The top row shows the energy in the  dipole mode (quadratic sum of the $m=0$ and $m=\pm 1$ components). The second row shows the energy in the quadrupole ($m=0$, $m=\pm 1$, and $m=\pm 2$) mode. The third row shows the energy in the octupole ($m=0$, $m=\pm 1$, $m=\pm 2$, and $m=\pm 3$) mode. The bottom row displays the monthly sunspot number during this time, taken from WDC-SILSO.}
   \label{fig:dipquadoct}
\end{figure}

\begin{figure}
 \centering
  \includegraphics[trim=0cm 0cm 0cm 0cm, clip, width=0.5\textwidth]{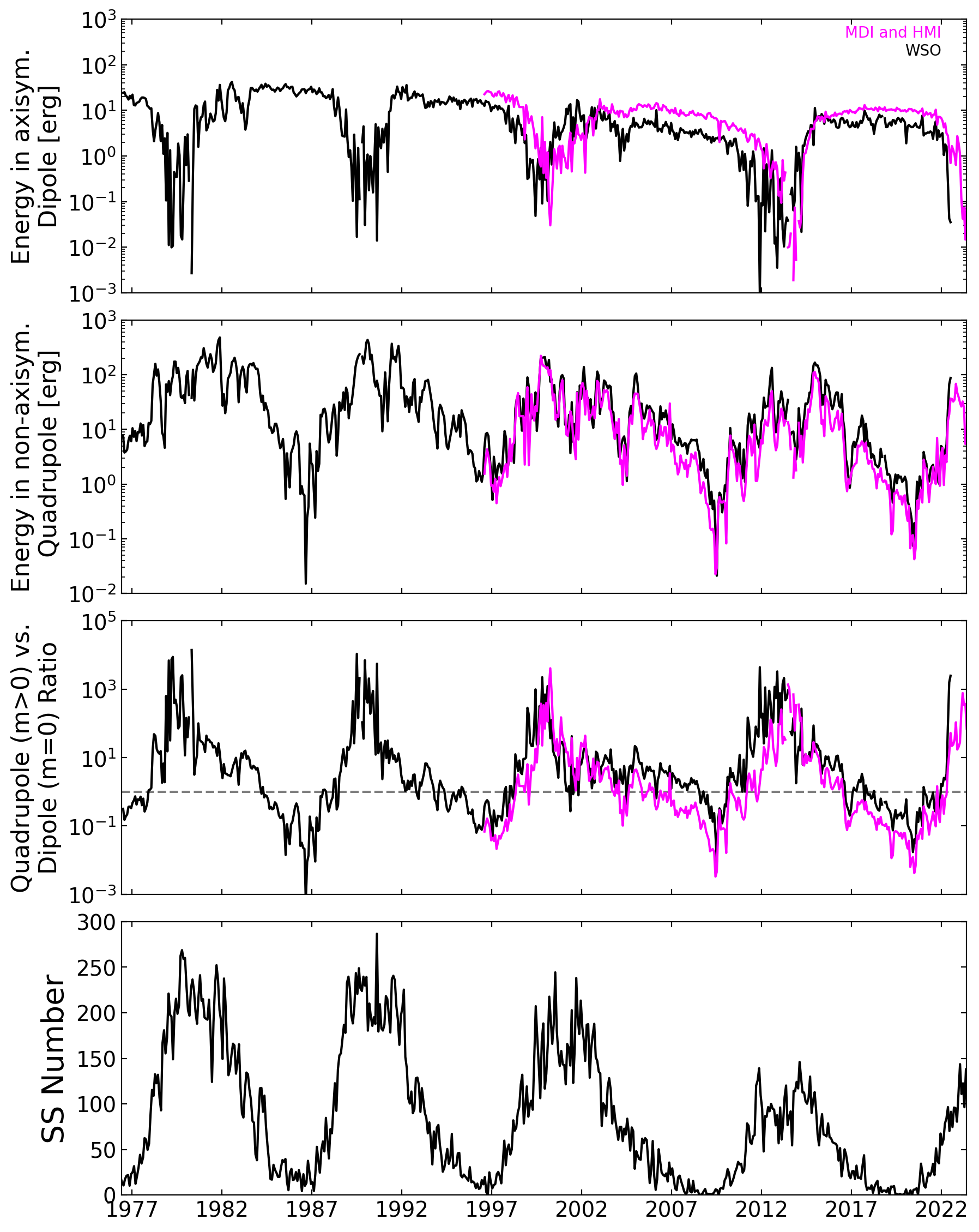}
   \caption{Comparison of energy in the axisymmetric dipole ($l=1$, $m=0$) and non-axisymmetric quadrupole components ($l=2$, $m>0$) versus solar cycle. The top row shows the energy in the axisymmetric dipole component. The second row shows the energy in the quadrupole $m=\pm 1$, and $m=\pm 2$ components quadratically summed. The third row shows the ratio of energy between the two energies, which a dashed horizontal line marking unity. The bottom row displays the monthly sunspot number during this time, taken from WDC-SILSO.}
   \label{fig:quaddipratio}
\end{figure}

\begin{figure*}
 \centering
  \includegraphics[trim=0cm 0cm 0cm 0cm, clip, width=\textwidth]{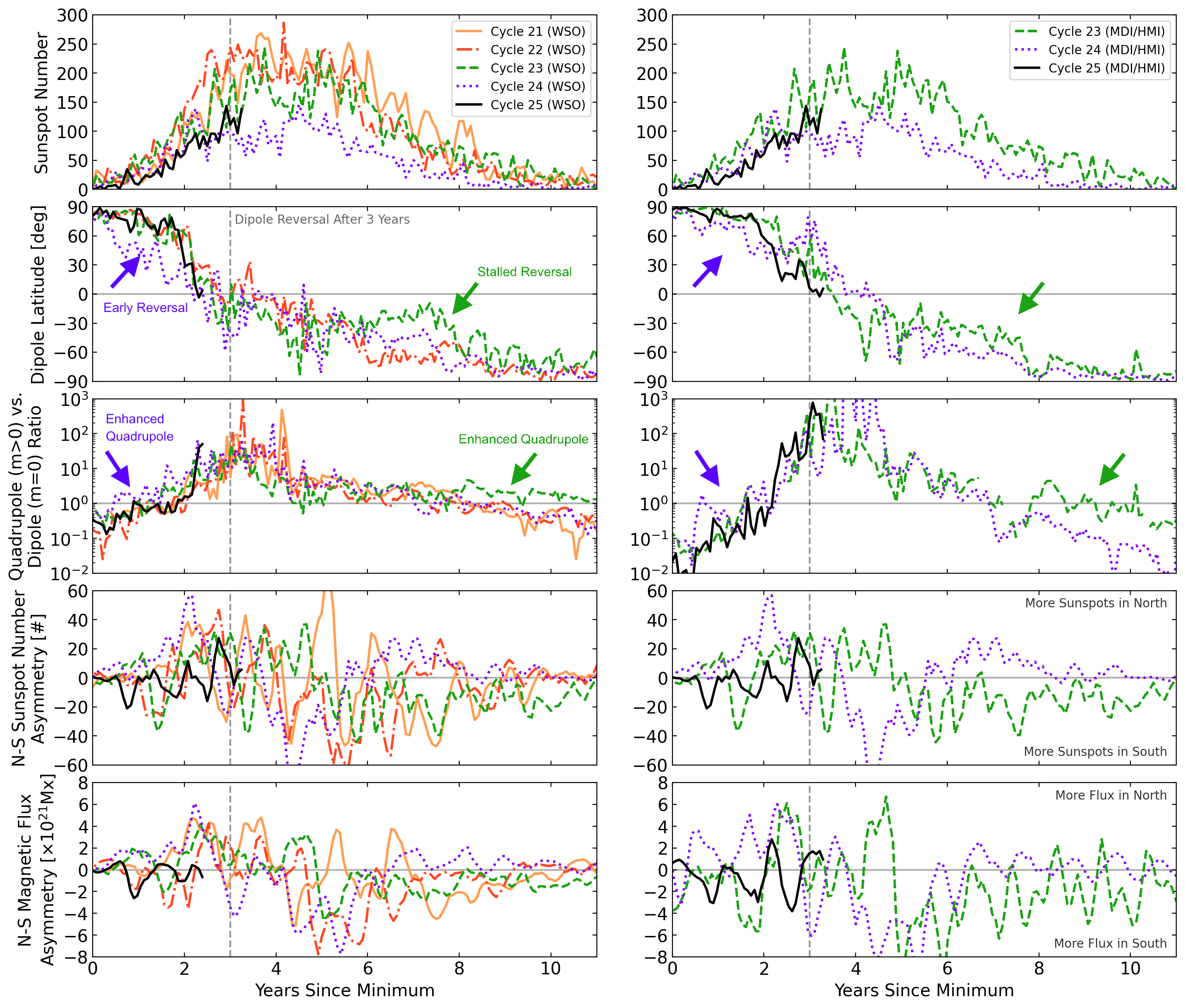}
   \caption{Solar cycle comparison of the monthly sunspot number, the dipole inclination angle, the ratio of non-axisymmetric quadrupole to axisymmetric dipole, and the hemispherical asymmetry in both sunspot number and unsigned magnetic flux, from both the WSO, and SOHO/MDI \& SDO/HMI timeseries. Each cycle is given a different line colour and style, cycle 25 is highlighted with a solid black line. Quantities are plotted with respect to the smoothed sunspot minima time of each cycle..}
   \label{fig:cycle_dipquadasym}
\end{figure*}

\begin{figure*}
 \centering
  \includegraphics[trim=0cm 0cm 0cm 0cm, clip, width=\textwidth]{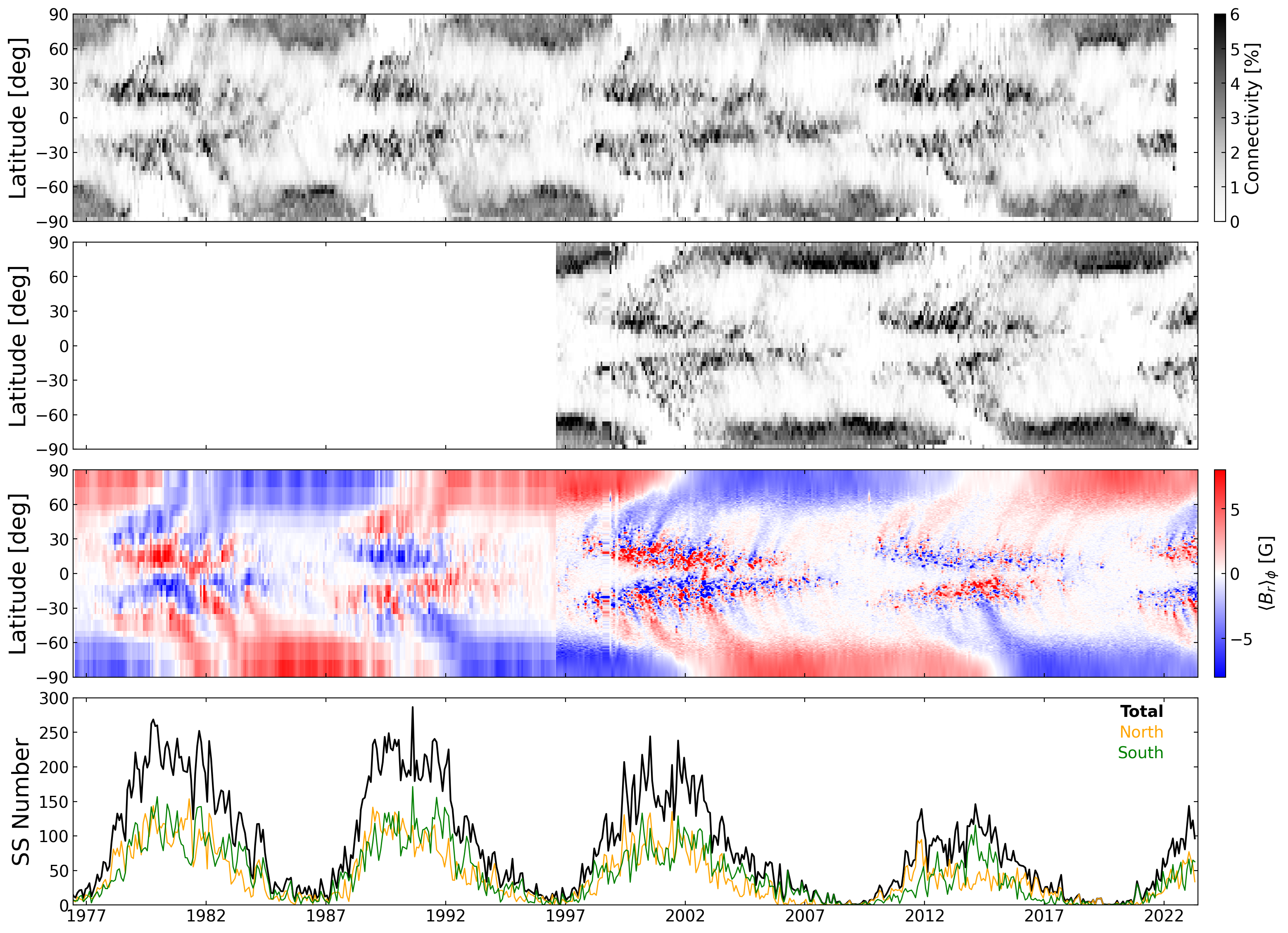}
   \caption{Solar cycle evolution of the solar wind footpoints. The top two panels show the fraction of open field lines, in each Carrington rotation, that trace down through the PFSS from a uniform distribution in latitude-longitude at the source surface to a given latitude bin at the surface (width $\sim 2^{\circ}$) using WSO magnetograms (1976-2022) and SOHO/MDI \& SDO/HMI magnetograms (1996-present), respectively (the longitudinal information is ignored, and the connectivity sums to 100\% for each Carrington rotation). The third panel displays a combined radial magnetic field butterfly diagram, in which the pole-ward surges of magnetic flux are easily distinguished. The bottom row displays the monthly sunspot number during this time, taken from WDC-SILSO, along with the contributions from the northern and southern hemispheres individually.}
   \label{fig:connect}
\end{figure*}

\section{Component analysis}\label{results1}
Each magnetogram in our timeseries, provided on a latitude $\theta$ versus Carrington longitude $\phi$ grid, was decomposed into spherical harmonics\footnote{To compute these coefficients the pySHTOOLS python package is used, which provides access to the Fortran-95 SHTOOLS library.} $Y_{lm}=c_{lm}P_{lm}(\cos\theta)e^{im\phi}$, with degree $l$ and order $m$, based on the legendre polynomial functions $P_{lm}(\cos\theta)$. The radial field is written as,
\begin{equation}
    B_r(\theta,\phi) = \sum_{l=1}^{l_{max}}\sum_{m=-l}^{l} B_{lm} Y_{lm}(\theta,\phi),
\end{equation}
with the normalisation, 
\begin{eqnarray}
c_{lm}=\sqrt{\frac{2l+1}{4\pi}\frac{(l-m)!}{(l+m)!}},
\label{clm}
\end{eqnarray} 
where the coefficients $B_{lm}$ denote the strength of each component. The maximum spherical harmonic degree recovered $l_{max}$ was 30 for the WSO magnetograms and 90 for the SOHO/MDI \& SDO/HMI magnetograms, based on the available resolution. The variation of the dipolar ($l=1$), quadrupolar ($l=2$), and octupolar ($l=3$) energies recovered from the magnetogram timeseries are shown in Figure \ref{fig:dipquadoct}, along with the monthly sunspot number\footnote{Data accessed May 2023: https://www.sidc.be/SILSO/datafiles}. Differences between the ground-based and space-based timeseries primarily result from the polar field strengths, with the combined SOHO/MDI \& SDO/HMI timeseries including a polar field correction. Supported by the field strengths of the dipole and octupole having a larger discrepancy between the two timeseries than the quadrupole during solar minima. A more detailed breakdown of these components, along with the higher order modes, is included in Appendix \ref{ap2}.

\subsection{Dipole axis evolution}

As the dipole mode has the slowest radial decay (in comparison to the quadrupole, octupole, etc), it has a significant role in shaping the coronal field. At large distances, the coronal field often appears dipolar, despite the complexity visible in the low-corona \citep[seen during some solar eclipses][]{mikic2018predicting}. From the spherical harmonic decomposition of the magnetogram timeseries, the evolution of the dipole axis is shown in the third and fourth panels of Figure \ref{fig:dipangle}. Values for latitude and longitude are given for the positive pole. There are some differences between the two timeseries, but overall the agreement is good. The cyclic reversals of the large-scale magnetic field during the solar cycle are easily identifiable, with the dipole reversing quickly during the onset of the cycle, reaching a fully inclined position around three years after sunspot minimum. The dipole then slowly completes its reversal during the remainder of the  $\sim11$ year cycle, returning back to the same overall polarity after $\sim22$ years. 

The dipole reverses during activity maxima once the pre-existing polar fields have been cancelled by newly emerged flux transported towards the poles \citep{mordvinov2019evolution}. However, each reversal is not smooth. Several epochs of stalling are identified with black horizontal bars in Figure \ref{fig:dipangle}. In addition, during these stalling epochs, the dipole axis slowly drifts in Carrington longitude (shown by the evolving colour of the points). This drift relates either to the local rotation rate of the magnetic flux comprising the dipole mode, or the emergence/decay of strong active regions and ephemeral regions contributing to the dipole mode. A more systematic analysis is left for future works.


\subsection{Quadrupole versus dipole ratio}\label{quadrupole}

The ratio of quadrupole to dipole energy varies with solar activity \citep[see][]{derosa2012solar}, with the quadrupolar energy overwhelming the dipolar energy during solar maxima, and vice versa during solar minima. The time-evolution of the dipole and quadrupole energies derived from the magnetogram timeseries are shown in Figure \ref{fig:dipquadoct} (see also Appendix \ref{ap2}). Each mode has an axisymmetric component ($m=0$), along with non-axisymmetric component(s). For the quadrupole mode, the axisymmetric field strengths are significantly weaker than the non-axisymmetric components (see Figure \ref{fig:quadbreakdown}), unlike the dipole whose components have similar strengths but appear with a phase lag (see Figure \ref{fig:dipbreakdown}). As both dipolar and quadrupolar energies are largest during the maximum of activity, they are both sensitive to the strength of active regions. During solar minima, the dipolar energy scales with the polar field strengths. As the polar fields weaken with rising activity, the dipole then becomes more sensitive to the contributions from active regions. In contrast, the quadrupolar energy is mostly linked to the emergence of active regions. 

The ratio of non-axisymmetric quadrupolar ($m>0$) and axisymmetric dipolar ($m=0$) energies, therefore reflects the competition between the polar fields and active regions in sculpting the large-scale coronal field. The ratio between these two terms is shown in Figure \ref{fig:quaddipratio}. The selection of these components, in comparison to using the total dipolar and quadrupolar energies, as was done in \citet{derosa2012solar}, accentuates the underlying trend. A smaller quadrupolar to dipolar ratio signifies more open field concentrated at the rotational poles, and a larger value means more field emerging at active solar latitudes. This is discussed further in Section \ref{results3}.

\subsection{Cycle to cycle variation}

Figure \ref{fig:cycle_dipquadasym} shows the monthly sunspot number, dipole latitude, quadrupole to dipole ratio, along with the north-south asymmetry in monthly sunspot number and unsigned magnetic flux, for the WSO and SOHO/MDI \& SDO/HMI timeseries respectively. Each quantity is plotted in time with respect to the start of each solar cycle, as defined by the World Data Center SILSO\footnote{Data accessed May 2023: https://www.sidc.be/SILSO/cyclesmm}, at the Royal Observatory of Belgium (see Table \ref{table:cycles}). The dipole latitude follows whichever polarity is in the northern hemisphere at cycle minima, such that each reversal progresses from north to south. The reversal of the dipole mode progresses faster during the rising phase, with the dipole mode fully inclined around three years after sunspot minimum. Consequently, the quadrupole to dipole ratio, as defined here using the axisymmetric dipole, also changes regime around this time. Despite significant differences in solar cycle strengths (up to a factor of two in sunspot number), the evolution of the dipole inclination angle and quadrupole to dipole ratio is very similar from cycle to cycle. Notable exceptions are the declining phase of cycle 23 and the onset of cycle 24. In both cases, the dipole is more inclined than the other cycles, and the quadrupole to dipole ratio is elevated. 

From the lower panels of Figure \ref{fig:cycle_dipquadasym}, the north-south asymmetry in sunspot number and unsigned magnetic flux (however noisy) follows the same trend shown by the hemispherical lag times in Figure \ref{fig:cyclelags}, peaking for the northern hemisphere before the southern hemisphere. The asymmetric variation in sunspot number and unsigned flux are correlated, as they principally measure the same phenomena, except that the unsigned flux is also sensitive to the polar fields and ephemeral regions. The north-south asymmetry is generally limited to a difference of around 60 sunspots or $8\times 10^{21}$Mx of unsigned flux between hemispheres.

\section{Solar wind sources and coronal rotation}\label{results2}

The coronal magnetic field topology was reconstructed using a PFSS model \citep{altschuler1969magnetic, schrijver2003photospheric}. Coronal magnetic fields are computed efficiently due to the simplicity of the PFSS model, and so it can be applied to large datasets of magnetograms. The PFSS models were driven by the spherical harmonic coefficients extracted in Section \ref{results1} from the radial field at the photosphere. The magnetic field in the model was constrained to be current-free ($\nabla\times B = 0$) and purely radial at the source surface (fixed at $2.5R_{\odot}$ in this study). The coronal magnetic field is described by,
\begin{eqnarray}
B_r(r,\theta,\phi) = \sum_{l=1}^{l_{max}}\sum_{m=-l}^{l}\alpha_{lm}(r)Y_{lm}(\theta,\phi),\\
B_{\theta}(r,\theta,\phi) = \sum_{l=1}^{l_{max}}\sum_{m=-l}^{l}\beta_{lm}(r)Z_{lm}(\theta,\phi),\\
B_{\phi}(r,\theta,\phi) = \sum_{l=1}^{l_{max}}\sum_{m=-l}^{l}\beta_{lm}(r)X_{lm}(\theta,\phi),
\end{eqnarray}
with,
\begin{eqnarray}
Y_{lm}&=&c_{lm}P_{lm}(\cos\theta)e^{im\phi},\\ 
Z_{lm}&=&\frac{c_{lm}}{l+1} \frac{dP_{lm}(\cos\theta)}{d\theta} e^{im\phi}, \\ 
X_{lm}&=&\frac{c_{lm}}{l+1} P_{lm}(\cos\theta) \frac{im}{\sin\theta} e^{im\phi},
\end{eqnarray}
where $r$ denotes the radial distance from the origin, $\alpha_{lm}(r)$ and $\beta_{lm}(r)$ are functions denoting the radial dependence of each spherical harmonic component \citep[see][and reference therein]{finley2023accounting}, and the normalisation $c_{lm}$ follows equation (\ref{clm}). In this study, the coronal magnetic field was reconstructed using an $l_{max}$ of 30 in all the PFSS reconstructions.

\begin{figure*}
 \centering
  \includegraphics[trim=0cm 0cm 0cm 0cm, clip, width=\textwidth]{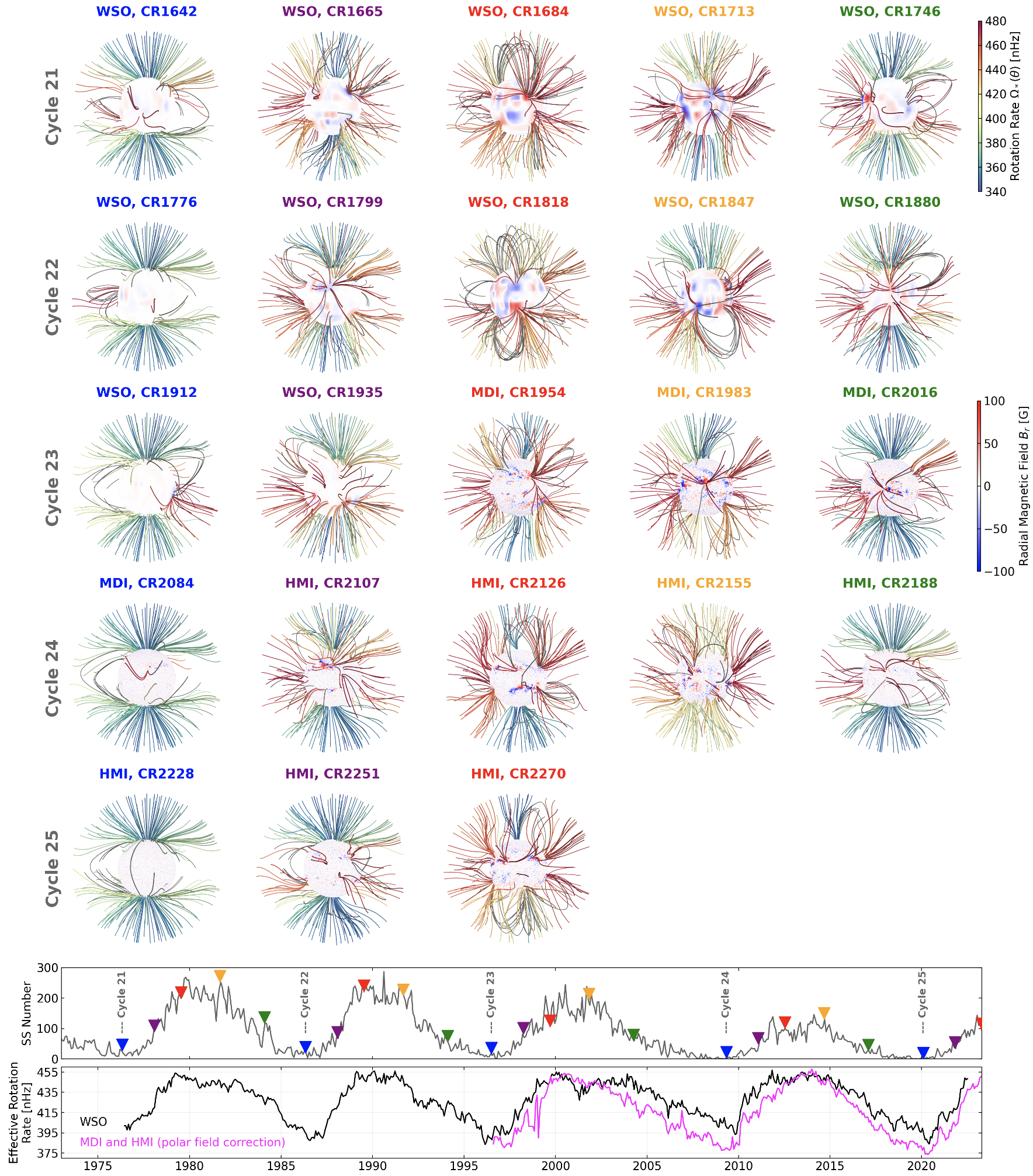}
   \caption{Snapshots of the coronal magnetic field at; 1) sunspot minimum (blue label), 2) after $\sim 1.5$ years (purple), 3) after $\sim 3$ years (red), 4) after $\sim 5$ years (orange), and 5) after $\sim 7.5$ years (green), for solar cycles 21-25 (cycle 25 is in progress). Each PFSS field line extrapolation is coloured by the photospheric rotation rate at its footpoint. The surface of each model is coloured by the corresponding radial magnetic field strength. The source of each magnetogram, along with the Carrington rotation number is given in the labels. The monthly sunspot number is shown below, with the time of each snapshot identified by a coloured marker. The bottom row displays the effective rotation rate of the open magnetic field at the source surface (weighted by $\sin\theta$), derived from a timeseries of PFSS models using the WSO (black) or SOHO/MDI \& SDO/HMI (magenta) Carrington magnetograms.}
   \label{fig:PFSS}
\end{figure*}

\subsection{Latitudinal connectivity}

PFSS models were computed for every magnetogram in each timeseries. The distribution of solar wind source latitudes was evaluated by tracing field lines down from the source surface to the photosphere. The resulting distribution was summed in longitude and stacked in time in order to produce the connectivity butterfly diagrams in Figure \ref{fig:connect}. The top panel contains the WSO timeseries, and the second panel the SOHO/MDI \& SDO/HMI timeseries. Differences between the two timeseries arise primarily from the relative resolutions of the underlying magnetogram timeseries. The WSO connectivity butterfly diagram is less sharp, but in general shows the same patterns and trends as the SOHO/MDI \& SDO/HMI timeseries. 

During each activity cycle, the sources of the open coronal magnetic field, and by extension the solar wind, evolve from the polar field at minimum to the active latitudes as solar activity increases \citep[explored in][]{stansby2021active}. The emergence of active regions distorts the coronal field and provides new source regions, be it the active regions themselves, equatorial coronal holes, or ephemeral regions. Easily identified are the pole-ward surges of magnetic flux \citep[see discussion in][]{finley2023accounting}, which also account for a significant fraction of the open field during the decaying phase of the activity cycle, as the polar fields begin to regenerate.

\subsection{Coronal rotation}

From the cyclic variation of footpoint latitudes identified with the PFSS modelling, the impact on coronal rotation was estimated by extrapolating the photospheric rotation rate along open magnetic field lines. The conservation of effective rotation rate is assumed, i.e. the balance between rotational flows and magnetic stresses in the field lines is maintained with height, as in \citet{finley2023accounting}. This does not account for the torques exerted between neighbouring field lines anchored at different latitudes, or time-dependent changes to the coronal field. Despite these caveats, the mean effective rotation rate is thought to be a reasonable constraint on the solid-body rotation rate required to match the angular momentum-loss rate of the solar wind for a given epoch \citep[see][]{ireland2022effect}. The Sun's photospheric rotation rate was parameterised as,
\begin{equation}
    \Omega_*(\theta) = \Omega_{eq}+\alpha_2\cos^2\theta+\alpha_4\cos^4\theta,
    \label{omega_*}
\end{equation}
where $\Omega_{eq}$ is the equatorial rotation rate, and the values of $\alpha_2$ and $\alpha_4$ describe the north-south symmetric differential rotation profile. Values of $\Omega_{eq}=472.6$ nHz, $\alpha_2=-73.9$ nHz, and $\alpha_4=-52.1$ nHz were adopted from \citet{snodgrass1983magnetic}, which are consistent with \citet{finley2023accounting}. A version of Figure \ref{fig:connect} colouring the connectivity butterfly diagram with the photospheric rotation rate is available in Appendix \ref{ap3}.

Snapshots from the PFSS modelling are shown in Figure \ref{fig:PFSS}, with open field lines coloured by their photospheric rotation rates. For each activity cycle, five PFSS models have been selected, one at minimum of activity, then after $\sim 1.5$ years (rising phase), $\sim 3$ years (reversal), $\sim 5$ years (high activity), and finally $\sim 7.5$ years (declining phase). These epochs are indicated with coloured arrows in the lower panel with the monthly sunspot number. During each cycle, the overall trend is the same. As activity increases, field lines connect to more rapidly rotating latitudes at the photosphere, and so the open field lines are more frequently red-orange tones, in comparison to the slowly rotating blue-green of solar minima. The declining phase of each cycle typically contains a few decaying active regions which maintain areas of enhanced rotation, mostly absent at solar minima.

\begin{figure}
 \centering
  \includegraphics[trim=0cm 0cm 0cm 0cm, clip, width=0.5\textwidth]{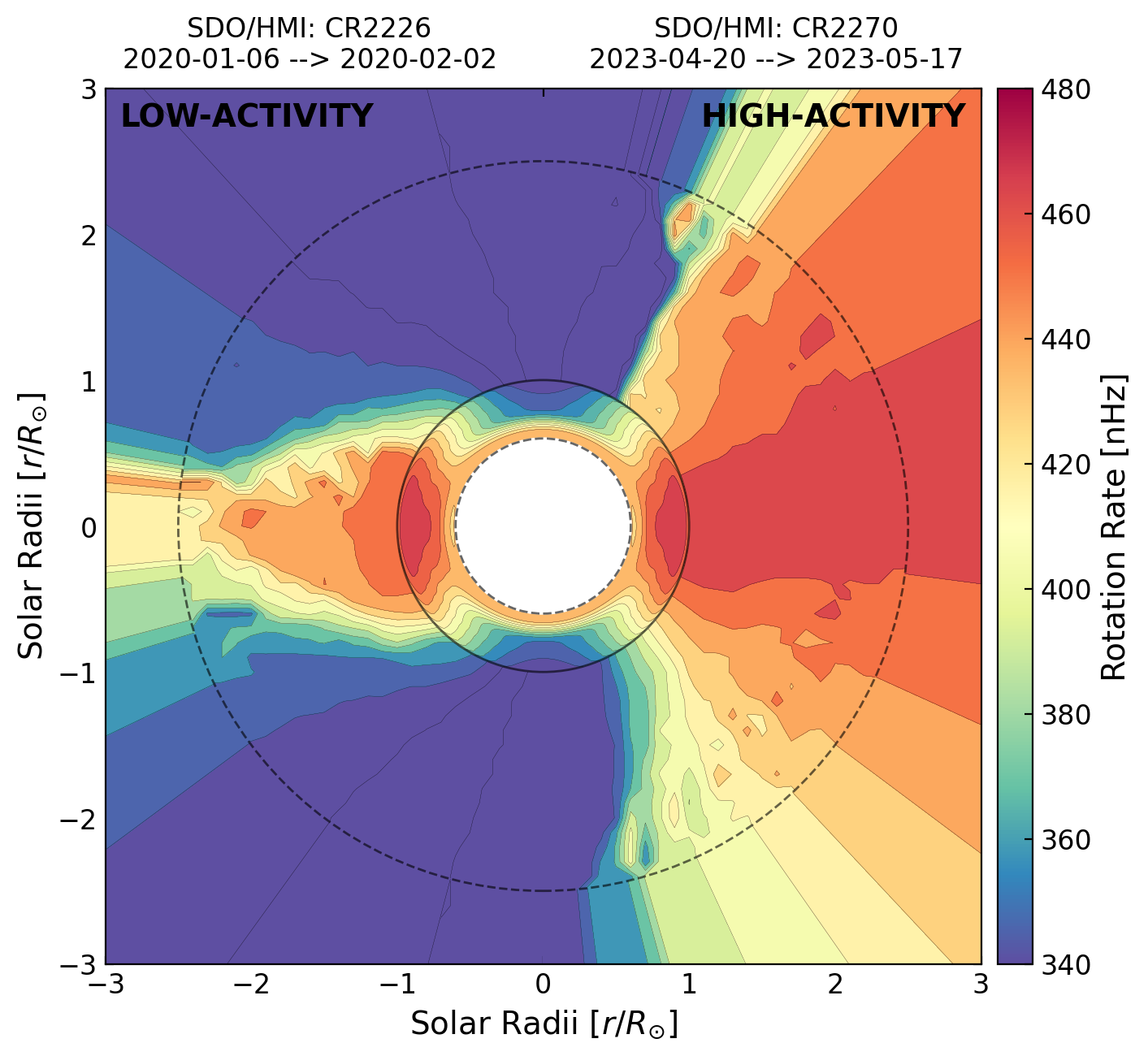}
   \caption{Comparison of azimuthally averaged footpoint rotation rates in the corona from low solar activity to high solar activity, left and right respectively. The solid black circle distinguishes the solar photosphere, below which the internal rotation profile derived from helioseismology is shown \citep[see][]{larson2018global}. The outer dashed line shows the source surface of the underlying PFSS model, beyond which the magnetic field is radial.}
   \label{fig:min_max}
\end{figure}

\begin{figure*}
 \centering
  \includegraphics[trim=0cm 0cm 0cm 0cm, clip, width=\textwidth]{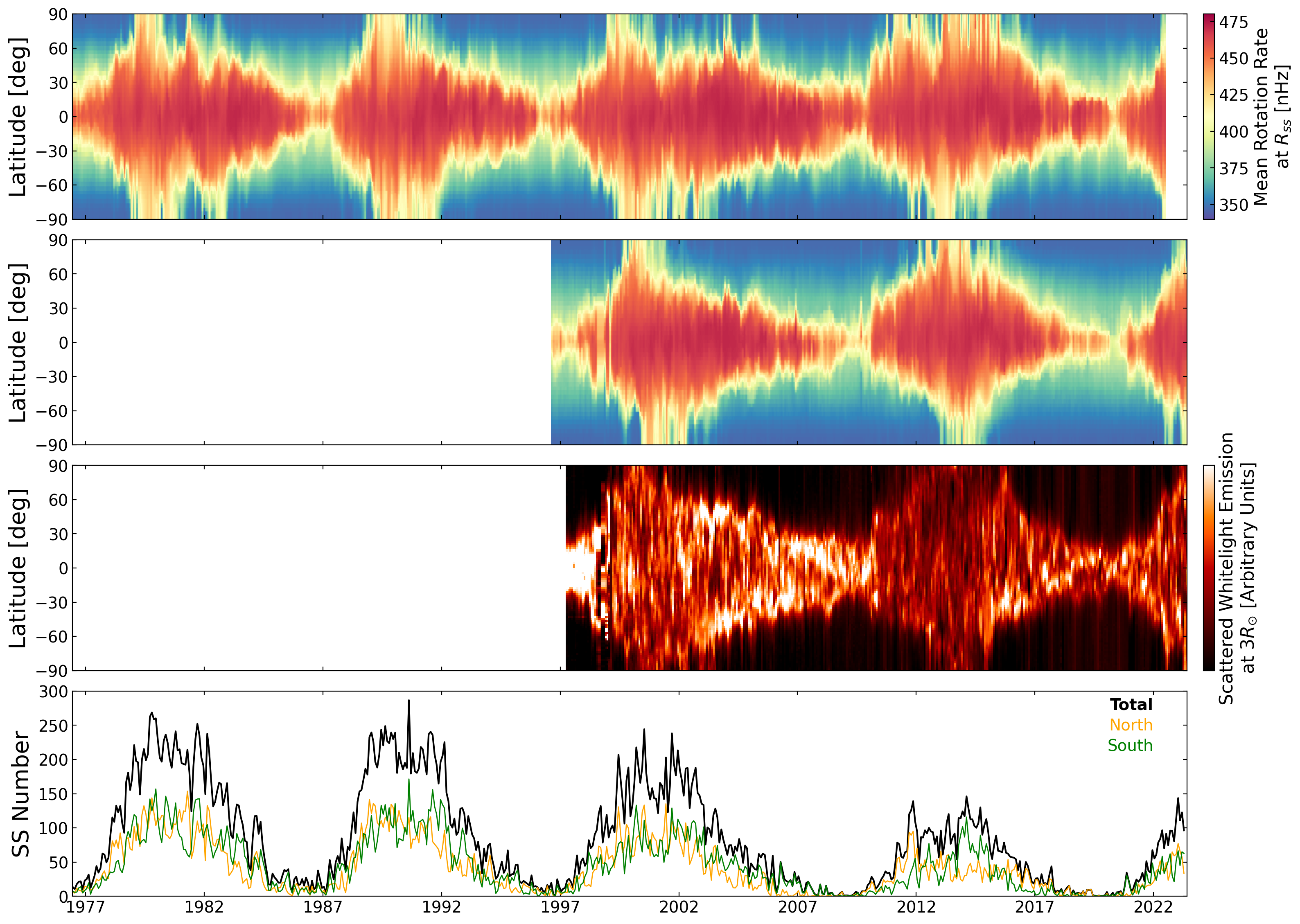}
   \caption{Solar cycle evolution of coronal rotation. The top two panels show the azimuthally averaged rotation rate at the source surface of the PFSS models using WSO magnetograms (1976-2022) and SOHO/MDI \& SDO/HMI magnetograms (1996-present), respectively. The third panel displays the average brightness of scattered white light from streamers at three solar radii observed by SOHO/LASCO-C2. The bottom row displays the monthly sunspot number during this time, taken from WDC-SILSO, along with the contributions from the northern and southern hemispheres individually.}
   \label{fig:lasco}
\end{figure*}

\begin{figure*}
 \centering
  \includegraphics[trim=0cm 0cm 0cm 0cm, clip, width=\textwidth]{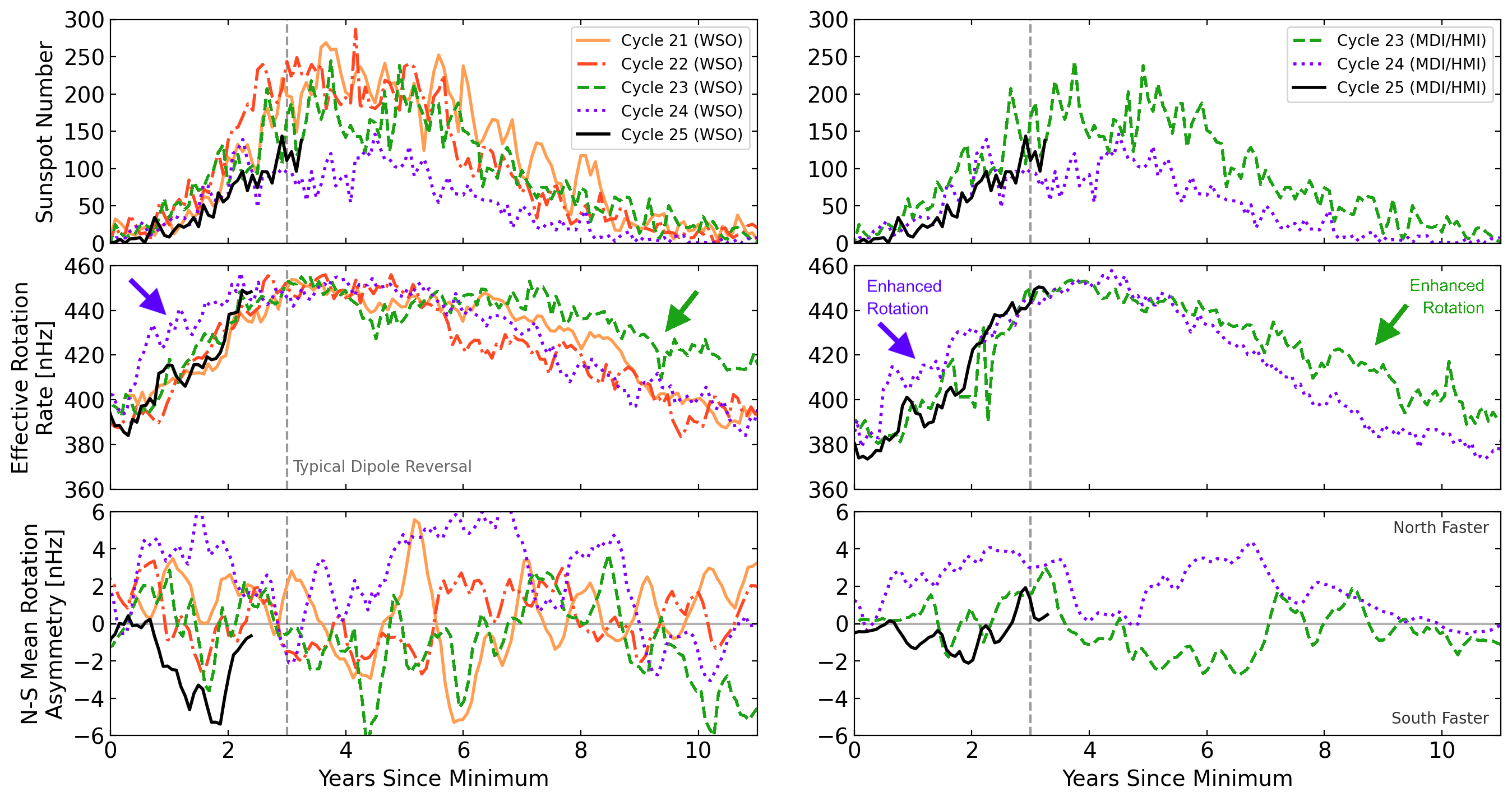}
   \caption{Solar cycle comparison of the monthly sunspot number, the effective rotation rate, and the hemispherical asymmetry in rotation, from both the WSO, and SOHO/MDI \& SDO/HMI timeseries. Each cycle is given a different line colour and style, cycle 25 is highlighted with a solid black line. Quantities are plotted with respect to the smoothed sunspot minima time of each cycle.}
   \label{fig:cycles}
\end{figure*}

The azimuthally averaged rotation rate in the corona was extracted from the PFSS models using the rotation rates traced along open magnetic field lines. This avoids the degeneracy from closed field lines whose footpoints are anchored at different latitudes. If no values were available for a given latitude and radius, i.e the location contained only closed field, a value was linearly interpolated from neighbouring regions. Figure \ref{fig:min_max} contrasts the result of this process for two PFSS models, one from solar minimum and one from solar maximum (both from the SDO/HMI magnetograms timeseries). During solar minima, the open field lines follow the axisymmetric dipole, with their footpoints anchored towards the rotational poles (reflected in their rotation rates). As activity increases, the solar dipole becomes highly inclined and the corona has more open field lines anchored to the faster equatorial and low-latitude regions. When the solar dipole is fully inclined, as is the case in Figure \ref{fig:min_max}, the closed dipole loops tend to close-off the slowly rotating polar regions, and the majority of the open field originates from active latitudes rotating near the Carrington rotation rate.


In Figure \ref{fig:lasco}, the time-evolution of the azimuthally averaged rotation rate at the source surface is shown. This tracks the latitudinal evolution of the rotation rate anchoring the solar wind versus time. Both timeseries show that during low-activity periods, the majority of open magnetic field is anchored in the slowly rotating poles. Connectivity to faster equator-ward sources is limited during solar minima, but as the cycle advances the range of latitudes containing faster, Carrington-like, rotation increases dramatically. During solar maxima, the open magnetic field can originate primarily from low latitude sources and so most latitudes are driven by faster rotation. 

In the third panel of Figure \ref{fig:lasco}, the azimuthally averaged brightness of scattered whitelight from SOHO/LASCO-C2\footnote{Data accessed May 2023: https://ssa.esac.esa.int/ssa}, at a distance of three solar radii, is compared to the variation of the azimuthally averaged rotation rate at the source surface. The variation of brightness from LASCO-C2 follows the evolution of long-lived whitelight streamers in the corona. These streamers share a similar cyclic variation to the rotation rate, as the underlying magnetic structures are the same. As discussed in \citet{finley2023accounting}, the apparent rotation of the corona can be extracted from the time-evolution of streamers \citep[see work from][and others]{morgan2011rotation, edwards2022solar}. However the systematic correlation between the streamers and regions of fast rotation may bias the recovered rotation rates towards larger values. In this case, the observed rotation rate versus latitude profile would appear, on average, flatter than in reality, more consistent with a Carrington-like rotation of the corona.

The $\sin\theta$ weighted average of the rotation rate at the source surface was computed in order to quantify the time-varying effective rotation rate (or mean rotation rate of the corona). This quantity is shown in the bottom panel of Figure \ref{fig:PFSS} for both WSO and SOHO/MDI \& SDO/HMI timeseries. As the polar field strengths are stronger in the SOHO/MDI \& SDO/HMI magnetograms, due to their polar field correction, the coronal field during minima is more concentrated towards the rotational poles than for the WSO timeseries, and so the effective rotation rate is systematically smaller there. Using the range of values from the SOHO/MDI \& SDO/HMI timeseries, the effective rotation rate varies from around 380nHz (30.4 days) during solar minima to 450nHz (25.7 days) during solar maxima. The rise in effective rotation rate is systematically sharper than the decline, reflecting the abrupt onset of active region emergence, and dipole rotation, versus the slow decay of each activity cycle. Similar trends are observed in the rotation of the solar corona, extracted from SDO/AIA observations \citep[e.g.][]{sharma2020variation}. \citet{wu2023rotational} recently showed that the mean rotation rate of the transition region varies as a function of solar cycle, fastest during solar maxima then slowly declining towards solar minima.

\section{Discussion and conclusions}\label{results3}

The connectivity of the open magnetic field, derived from the PFSS modelling, follows a similar pattern across multiple cycles, driven by the competition of polar field strengths and active region emergence rates. The rotation rate imparted to the coronal magnetic field changes throughout each cycle, as the open field lines are anchored at different latitudes, and consequently different rotation rates. The long-term evolution of rotation is observed in coronal radio emission \citep{deng2020systematic}, sunspot number records, and other emission indices \citep{chandra2011periodicities}, as well as studies of the transition region \citep{zhang2023temporal}. The timeseries produced in this study spans over four solar cycles, making it possible to distinguish long-term, cycle to cycle, trends. 

\begin{figure*}
 \centering
  \includegraphics[trim=0cm 0cm 0cm 0cm, clip, width=\textwidth]{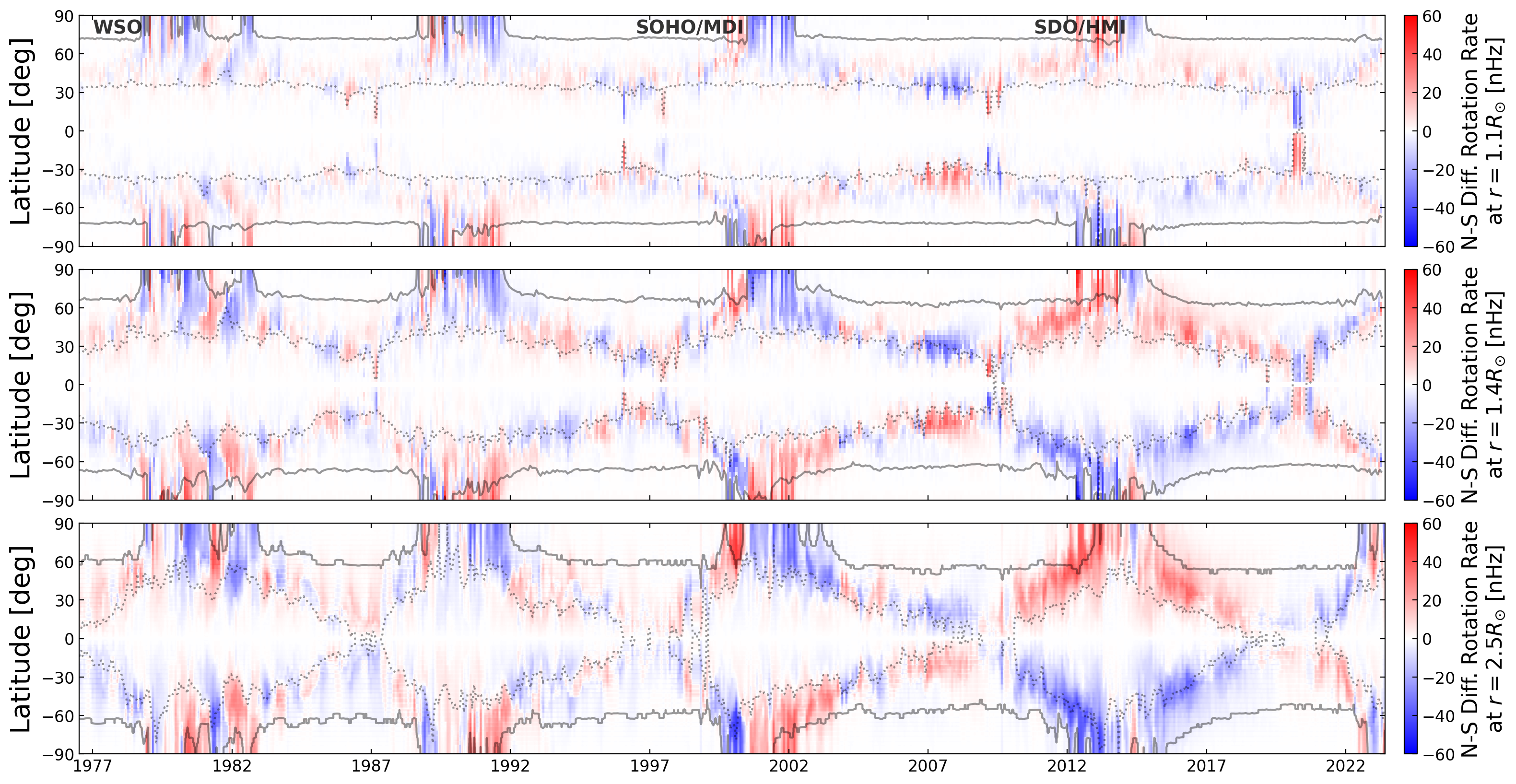}
   \caption{North-south asymmetry in coronal rotation at different altitudes (1.1, 1.4, and 2.5 solar radii). For each altitude, the solid and dotted lines are contours of rotation rate at 360 nHz and 440 nHz respectively. The colour indicates the difference in rotation rate between north and south at the same latitude, therefore values are anti-symmetric about the equator.}
   \label{fig:NSasym}
\end{figure*}

\subsection{Cycle to cycle variation in effective rotation rate}

The evolution of the effective rotation rate has strong similarities from cycle to cycle, visible in Figure \ref{fig:cycles}. The largest deviations from this occur during the declining phase of cycle 23 and the rising phase of cycle 24. Both corresponding to differences in the Sun's large-scale magnetic field between cycles. These deviations were previously highlighted in Figure \ref{fig:cycle_dipquadasym}.

Cycle 23 maintained an elevated effective rotation rate ($\sim 10 \%$ larger than average) throughout the declining phase, up until to the start of cycle 24. Figures \ref{fig:connect} and \ref{fig:lasco}, show cycle 23's extended cycle in both active region emergence (see magnetic butterfly diagram), and non-dipolar LASCO-C2 streamer structures, especially in comparison to cycle 24. During the declining phase, the quadrupole to dipole ratio remained elevated, unlike previous cycles whose ratio became dipole-dominated around 8 years after the minimum. In addition, the dipole mode was inclined to the rotation axis for much longer than during the other cycles. This lead to the open magnetic field being anchored to the more rapidly rotating low-latitude regions for a prolonged period, increasing the effective rotation rate. 

Cycle 24 was a weaker cycle with respect to the last $\sim 100$ years (see long-term comparison in \citealp{hathaway2015solar} and \citealp{nandy2021progress}, along with comparison of the upcoming cycle 25 in \citealp{upton2023solar}). This was likely driven by several large bipolar active regions that emerged at low-latitudes with the `wrong', or 'anti-Hale', north-south magnetic polarity with respect to Joy's law \citep{jiang2015cause}. Despite the strength of the cycle, the quadrupole to dipole ratio was larger during the rising phase than any of the other cycles in this study (perhaps a continuation of cycle 23). The dipole mode also started to reverse earlier (more evident in the WSO timeseries). As a consequence, the Sun's open magnetic field was connected to low-latitude sources earlier in cycle 24 and so the effective rotation rate was larger during the rising phase. 

\subsection{North-south Asymmetry}

Throughout the timeseries presented in this study, the latitudinal distribution of rotation at the source surface (see Figure \ref{fig:lasco}) is largely north-south symmetric, however the asymmetry is not zero. The mean rotation rate in each hemisphere differs by a few percent (see Figure \ref{fig:cycles}), driven by differences in the hemispherical flux emergence rates, and pole-ward surges of magnetic flux. This is most apparent during the rising and declining phases of the activity cycle, when the activity levels in each hemisphere can differ significantly (see bottom panels of Figure \ref{fig:cycle_dipquadasym}. Typically, this correlates with a rise in the axisymmetric quadrupolar component. 

The azimuthally averaged rotation rate at different altitudes were extracted from the PFSS model timeseries, as done in Figure \ref{fig:lasco} for the source surface. The northern and southern hemisphere rotation rates are then subtracted to examine the propagation of asymmetries. Figure \ref{fig:NSasym} shows the latitudinal asymmetry between hemispheres, at the altitudes of 1.1, 1.4, and 2.5 solar radii. The solid and dotted lines in each panel correspond to contours of rotation rate at 360 nHz, and 440 nHz, respectively. Lower down in the corona, these contours are similar to the imposed surface differential rotation rate, with deviations during maximum due to the polar field being cut-off by closed dipolar loops. Increasing in altitude, the latitudinal variation of rotation rate, visible in Figure \ref{fig:lasco}, appears.   

The largest local asymmetries are observed between the two contours, which corresponds to rotation rates slightly pole-ward of the active latitudes. As previously discussed in \citet{finley2023accounting}, asymmetry in the coronal field and rotation rate are often linked to the pole-ward surges of magnetic flux from decaying active regions. Accordingly, the asymmetries in Figure \ref{fig:NSasym} coincide with the pole-ward surges in Figure \ref{fig:connect}, and match with the mean hemispherical asymmetry in the rotation rate from Figure \ref{fig:cycles}. Interestingly, cycle 24 shows a very strong one-sided asymmetry, with the northern hemisphere being driven faster than the southern hemisphere throughout. The same trend is not visible in the observations for cycle 25 so far.


\subsection{Status of cycle 25}

Cycle 25 is well underway with activity proxies, like sunspot number, slightly elevated with respect to cycle 24. At the time of writing this paper, the dipole mode is fully-inclined to the rotation axis, and there is more energy in the quadrupole than the dipole (see Figure \ref{fig:cycle_dipquadasym}). The reversal of the dipole axis began along $\sim 190$ degrees Carrington longitude, until the reversal underwent a small stalling (annotated in Figure \ref{fig:dipangle}), during which time the dipole axis drifted to around $\sim 100$ degrees Carrington longitude, before advancing further towards the equator. After some initial north-south asymmetry in flux emergence, the open magnetic field footpoints have, as of May 2023, saturated around the active latitudes, leading to a coronal rotation rate around the Carrington value of 25.4 days. Cycle 25 is far more north-south symmetric than the previous four cycles examined in this study, following the weakening of the axisymmetric quadrupolar mode (see Figure \ref{fig:quadbreakdown} in Appendix \ref{ap2}). The north-south asymmetries in sunspot number, unsigned magnetic flux, and coronal rotation are compared with previous cycles in Figures \ref{fig:cycle_dipquadasym} and \ref{fig:cycles}, with cycle 25 generally more symmetric than cycle 24.

\subsection{Application to other Sun-like stars}

The strengths of the low-order modes, like the dipole, quadrupole, and octupole, can be recovered from the polarisation of magnetically sensitive spectral lines using the technique of Zeeman-Doppler imaging \citep[ZDI;][]{donati1989zeeman}. This has greatly increased our understanding of the magnetism of other Sun-like stars \citep{see2019estimating}, including their magnetic cycles \citep{jeffers2022crucial}. Stars like 61 Cyg A \citep{saikia2016solar} have been repeatedly observed with the ZDI technique, spanning more than a full chromospheric activity cycle, +revealing a similar evolution of the dipole inclination and quadrupole to dipole ratio to the Sun \citep{saikia2020solar}. As these low-order modes are more accessible, they could be leveraged in order to characterise the susceptibility of stellar coronal rotation to the differential rotation rate in the photosphere. This is important as differential rotation was recently proposed by \citet{tokuno2023transition} in order to explain the phenomenon of 'weakened magnetic braking' \citep{van2016weakened} in old-age low-mass stars like the Sun. Yet, the correct rotation rate to apply in rotation-evolution models \citep[e.g.][]{matt2015mass} when computing the angular momentum-loss rate is unconstrained. A characterisation of the average dipole tilt, and ratio of quadrupolar to dipolar energies for the stars already observed using ZDI, along with knowledge of their surface differential rotation rates (including the possibility of anti-solar differential rotation at late-ages e.g. \citealp{brun2022powering}, \citealp{noraz2022hunting}), has the potential to identify systematic trends in the effective rotation rate, needed for modelling the winds of these stars, and the subsequent loss of angular momentum responsible for their rotational-evolution \citep{ahuir2020stellar}.

\begin{acknowledgements}
This research has received funding from the European Research Council (ERC) under the European Union’s Horizon 2020 research and innovation programme (grant agreement No 810218 WHOLESUN), in addition to funding by the Centre National d'Etudes Spatiales (CNES) Solar Orbiter, and the Institut National des Sciences de l'Univers (INSU) via the Programme National Soleil-Terre (PNST).
Stanford University operates WSO with funding provided by the National Science Foundation with Grant No 1836370.
Data supplied courtesy of the SDO/HMI and SDO/AIA consortia. SDO is the first mission to be launched for NASA's Living With a Star (LWS) Program.
The sunspot number used in this work are from WDC-SILSO, Royal Observatory of Belgium, Brussels. 
Data manipulation was performed using the numpy \citep{2020NumPy-Array}, scipy \citep{2020SciPy-NMeth}, and pySHTOOLS \citep{wieczorek2018shtools} python packages.
Figures in this work are produced using the python package matplotlib \citep{hunter2007matplotlib}.
\end{acknowledgements}

%
%

\bibliographystyle{yahapj}
\bibliography{adam}

\begin{appendix}
\section{Comparison of WSO, SOHO/MDI, and SDO/HMI synoptic magnetograms} \label{ap1}

For the period of overlap between WSO, SOHO/MDI, and SDO/HMI (CR 2100 - CR 2107), where Carrington magnetograms are available from each observatory, the mean strength of the dipole, quadrupole, and octupole modes are tabulated in Table \ref{table:ap1}. These components are the most significant in shaping the large-scale coronal field, which is the focus of this study. In order to insure consistency between the three timeseries, the factor required to equate all observations to the strength of SDO/HMI was estimated. The mode strengths from SDO/HMI are on average 3.5 bigger than WSO, and so throughout this study WSO magnetograms are mutliplied by this factor. The mode strengths from SDO/HMI are 0.85 smaller, than SOHO/MDI, and so similarly SOHO/MDI magnetograms are multiplied by this factor. From Table \ref{table:ap1} however, there is a spread in values which varies from CR to CR. It is also clear that a multiplicative factor is not enough to restore the discrepancy between the ground-based WSO and the two space-based instruments. The field strengths may instead require a latitudinally varying factor, as the polar field strengths are challenging to recover. This primarily influences the odd $l$ modes, who directly reconstruct the anti-symmetric polar fields.

\begin{table*}
\caption{Dipole, Quadrupole, and Octupole strengths recovered from concurrent WSO, MDI, and HMI magnetograms, along with a ratio of their averaged strength.}
\label{table:ap1}
\centering
\begin{tabular}{c | c c | ccc ccc ccc | c} 
\hline\hline
CR No. & Start Date & End Date  & \multicolumn{3}{c}{Dipole [G]}      & \multicolumn{3}{c}{Quadrupole [G]}   &  \multicolumn{3}{c}{Octupole [G]}    & Av. Ratio\\
\cmidrule(rl){4-6} \cmidrule(rl){7-9} \cmidrule(rl){10-12}
     &            &           & WSO & MDI & HMI  & WSO & MDI & HMI   & WSO & MDI & HMI & WSO:MDI:HMI\\
\hline
2100 & 2010/08/09 & 2010/08/09 & 0.4 & 2.1 & 1.8 & 0.6 & 2.2 & 1.5 & 0.3 & 1.8 & 1.7 & 3.9:0.8:1.0 \\
2101 & 2010/09/05 & 2010/09/05 & 0.6 & 1.5 & 1.6 & 0.6 & 1.5 & 1.0 & 0.4 & 1.2 & 0.7 & 1.9:0.8:1.0 \\
2102 & 2010/10/03 & 2010/10/03 & 0.6 & 2.1 & 2.0 & 0.3 & 1.1 & 0.9 & 0.5 & 1.0 & 1.1 & 3.0:0.9:1.0 \\
2103 & 2010/10/30 & 2010/10/30 & 0.5 & 2.3 & 2.0 & 0.3 & 1.1 & 0.8 & 0.7 & 3.4 & 2.4 & 3.5:0.8:1.0 \\
2104 & 2010/11/26 & 2010/11/26 & 0.2 & 2.4 & 2.1 & 0.5 & 0.8 & 1.1 & 1.1 & 3.9 & 2.5 & 4.5:1.0:1.0 \\
2105 & 2010/12/24 & 2010/12/24 & 0.6 & 2.0 & 2.2 & 0.5 & 1.3 & 1.2 & 1.0 & 1.7 & 2.2 & 2.8:1.1:1.0 \\
2106 & 2011/01/20 & 2011/01/20 & 0.7 & 2.2 & 1.8 & 0.6 & 2.0 & 1.8 & 0.8 & 2.1 & 1.4 & 2.4:0.8:1.0 \\
2107 & 2011/02/16 & 2011/02/16 & 0.8 & 2.2 & 2.0 & 1.0 & 2.8 & 2.1 & 1.1 & 4.1 & 3.2 & 2.5:0.8:1.0 \\

\hline
\end{tabular}
\end{table*}

\section{Further component analysis} \label{ap2}

From the spherical harmonic decomposition of the magnetograms in Section \ref{results1}, the distribution of mode strengths during the last four cycles was examined. The lowest order modes (dipole, quadrupole, and octupole) significantly influence the large-scale structure in the corona. It is also interesting to examine trends between low and high order modes, plus odd and even families.

\subsection{Dipole mode}

\begin{figure}
 \centering
  \includegraphics[trim=0cm 0cm 0cm 0cm, clip, width=0.5\textwidth]{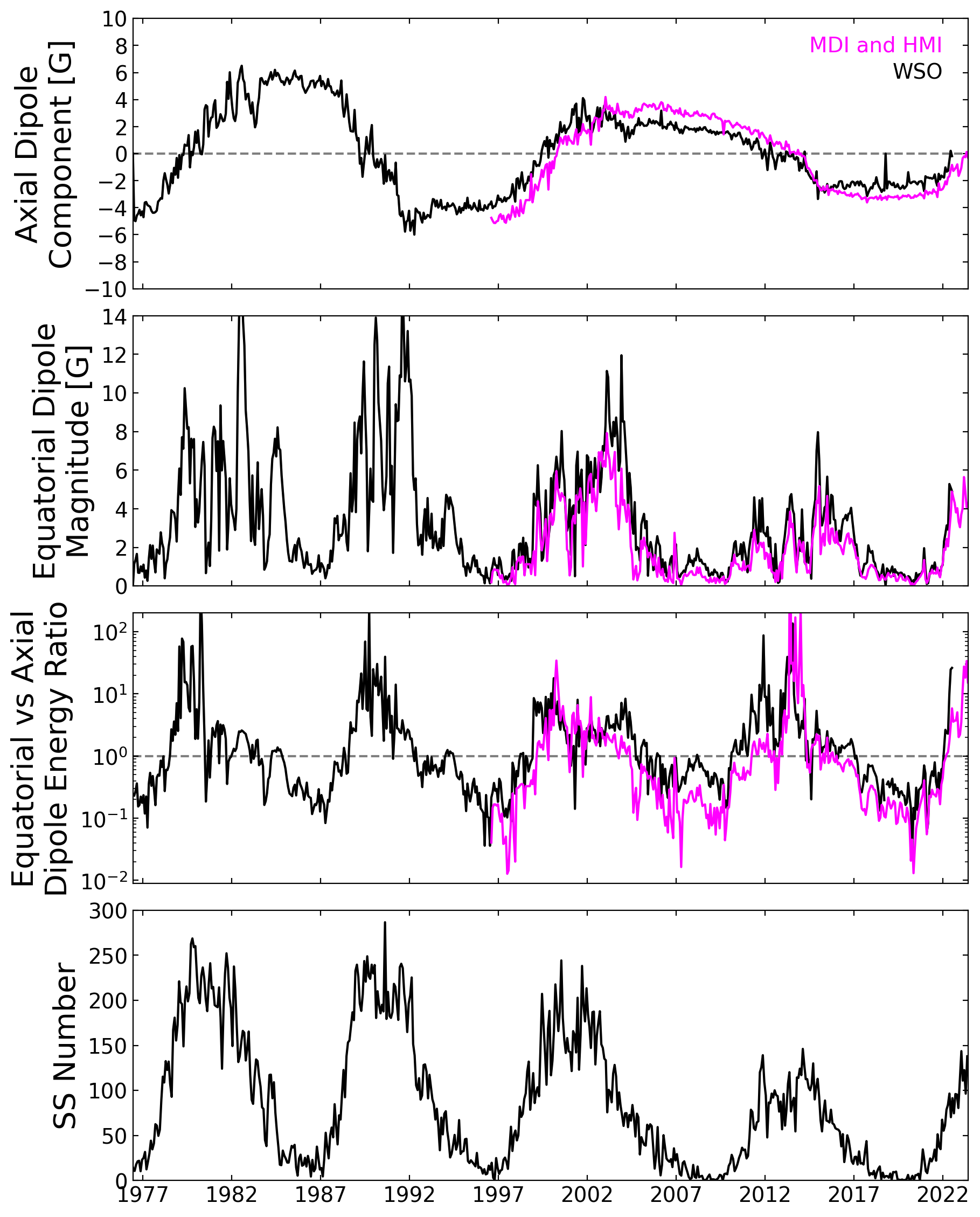}
   \caption{Evolution of the dipole ($l=1$) mode of the Sun's radial magnetic field. The first row shows the variation of the axial ($m=0$) component of the field. The second row shows the variation of the equatorial ($m=\pm 1$) component of the field. The third row shows the ratio of energy between the equatorial and axial dipole components. The bottom row displays the monthly sunspot number during this time, taken from WDC-SILSO.}
   \label{fig:dipbreakdown}
\end{figure}

The dipole mode undergoes a complete reversal each solar cycle, and so magnetic energy varies between the axisymmetric (axial) $m=0$ component, and the non-axisymmetric (equatorial) $m=\pm 1$ component. Figure \ref{fig:dipbreakdown} shows the strength of the axial and equatorial components of the dipole retrieved from the magnetogram timeseries, along with their ratio in the third panel. The equatorial component follows the sunspot number in the bottom row very closely, due to the sensitivity of this mode to active region emergence. The overall energy of the dipole mode has been declining during the last four cycles. At present, the axial component is reversing sign and the equatorial component is slightly larger than the rising phase of cycle 24. 

\subsection{Quadrupole mode}

The quadrupole field strength is mostly contained in the non-axisymmeric $m>0$ components, which follow the sunspot number, and are linked to active region emergence. However, despite the weakness of the axisymmetric $m=0$ component, enhancements in this component drive north-south asymmetries in the coronal field and are often associated with pole-ward surges of magnetic flux. An example of this is shown during cycle 24, when the axisymmetric mode grows due to the large southern pole-ward surge. This has implications for the asymmetry of the coronal field, which up until this surge had been controlled by the northern hemisphere (containing multiple smaller northward surges). This is reflected in the variation of the asymmetry in coronal rotation in Figure \ref{fig:cycles}.

\begin{figure}
 \centering
  \includegraphics[trim=0cm 0cm 0cm 0cm, clip, width=0.5\textwidth]{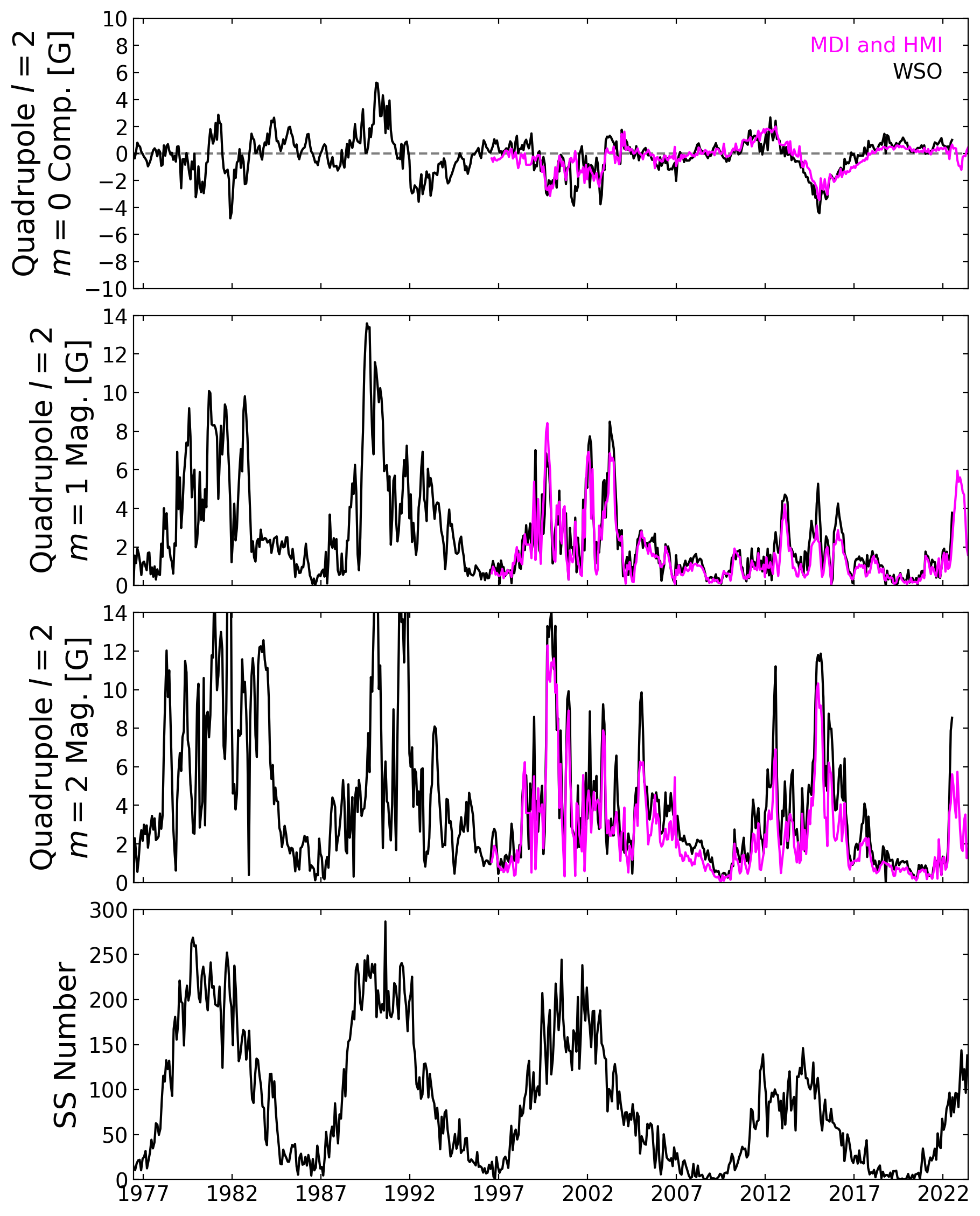}
   \caption{Evolution of the quadrupole ($l=2$) mode of the Sun's radial magnetic field. The first row shows the variation of the axial ($m=0$) component of the field. The following rows show the magnitudes of the higher order components. The bottom row displays the monthly sunspot number during this time, taken from WDC-SILSO.}
   \label{fig:quadbreakdown}
\end{figure}

\begin{figure}
 \centering
  \includegraphics[trim=0cm 0cm 0cm 0cm, clip, width=0.5\textwidth]{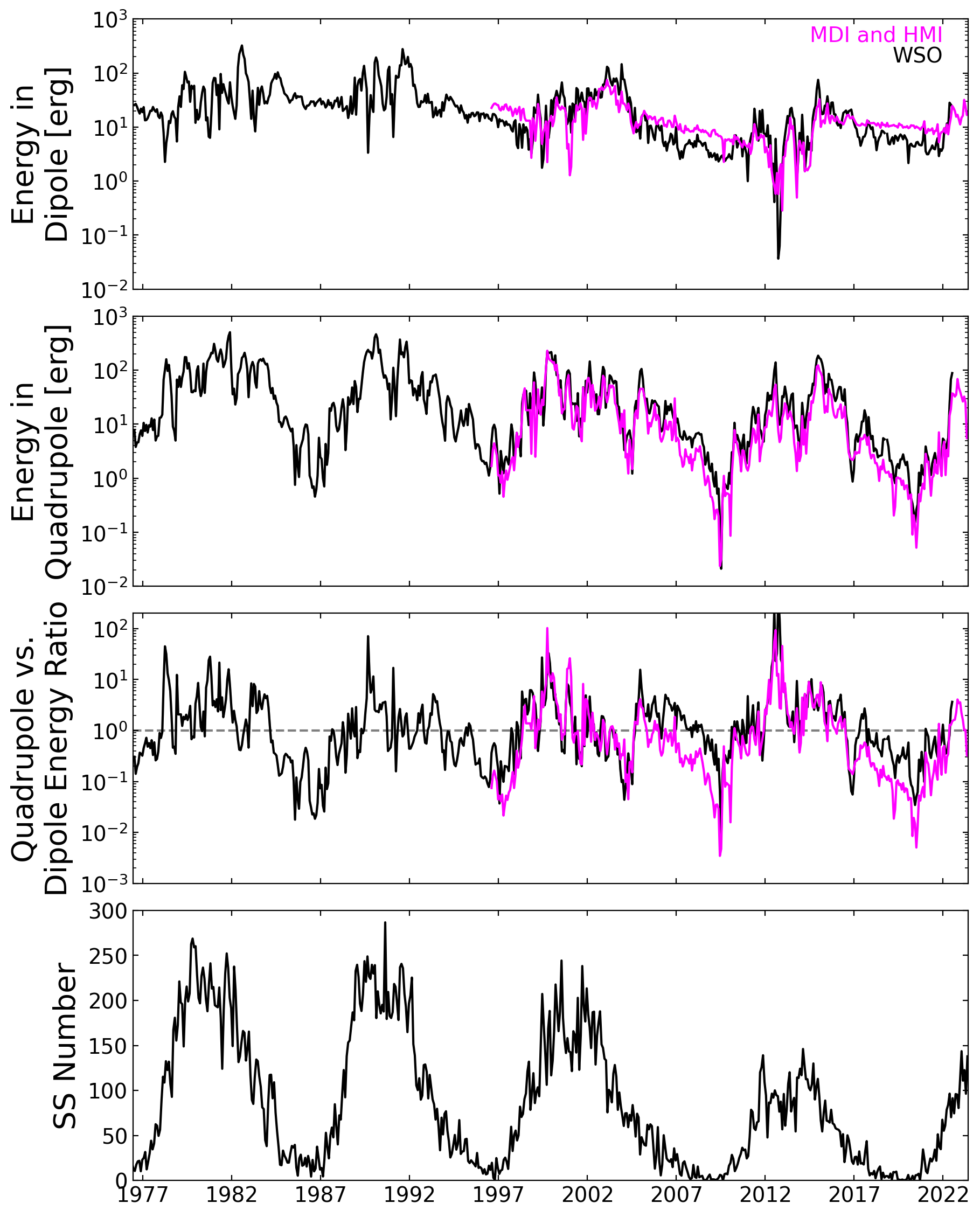}
   \caption{Comparison of energy in the dipole ($l=1$) and quadrupole ($l=2$) modes versus solar cycle. The top row shows the energy in the dipole mode (quadratic sum of the $m=0$ and $m=\pm 1$ components). The second row shows the energy in the quadrupole mode ($m=0$, $m=\pm 1$, and $m=\pm 2$ components). The third row shows the ratio of energy between the quadrupole and dipole modes. The bottom row displays the monthly sunspot number during this time, taken from WDC-SILSO.}
   \label{fig:quaddipratio_old}
\end{figure}

In addition to examining individual mode strengths, the strength of the dipolar and quadrupolar modes are compared, as in \citet{derosa2012solar} who noted that the solar cycle was well characterised by an oscillation of energy between the dipolar (anti-symmetric) and quadrupolar (symmetric) modes. Figure \ref{fig:quaddipratio_old}, shows the total dipolar and quadrupolar energy, along with their ratio. Another version of this figure, containing the ratio of axisymmetric dipole and non-axisymmetric quadrupolar energies, was produced for the main text of the paper (see Figure \ref{fig:quaddipratio}). This version accentuates the movement of energy from the polar fields to the active latitudes. In either case, at present the quadrupolar energy has surpassed that of the dipole, showing that the cycle 25 is well underway.

\subsection{Octupole mode}

The octupole component is an odd order $l$ mode meaning that, like the dipolar component, it is sensitive to the Sun's polar fields. The axisymmetric $m=0$ component has a very similar behaviour to the dipole $m=0$ mode. The non-axisymmetric $m>0$ modes are however much stronger than the equatorial dipole, and reflecting the increased amount of energy stored at this scale.

\begin{figure}
 \centering
  \includegraphics[trim=0cm 0cm 0cm 0cm, clip, width=0.5\textwidth]{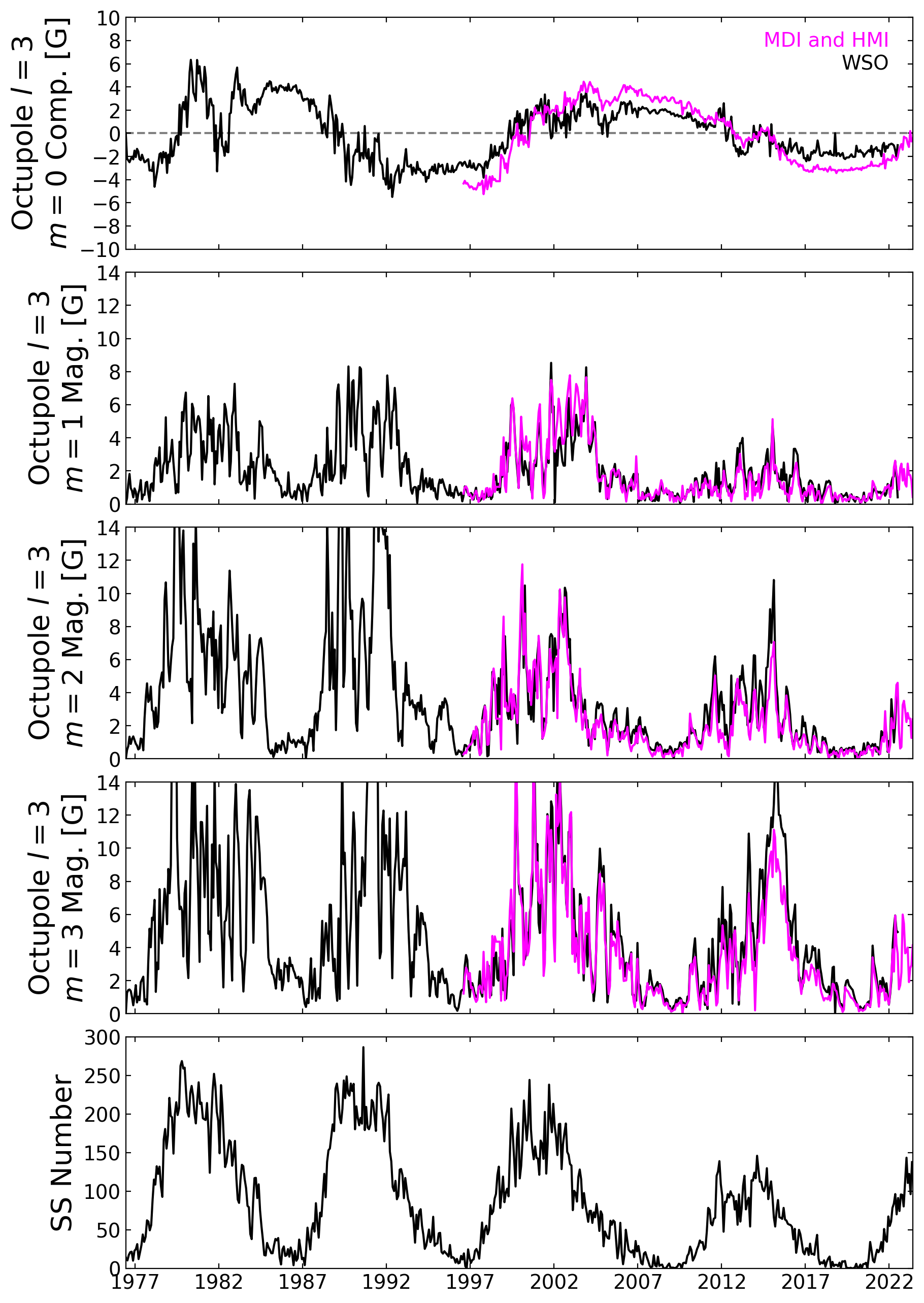}
   \caption{Evolution of the octupole ($l=3$) mode of the Sun's radial magnetic field. The first row shows the variation of the axial ($m=0$) component of the field. The following rows show the magnitudes of the higher order components. The bottom row displays the monthly sunspot number during this time, taken from WDC-SILSO.}
   \label{fig:octbreakdown}
\end{figure}

\subsection{Higher order modes}

\begin{figure}
 \centering
  \includegraphics[trim=0cm 0cm 0cm 0cm, clip, width=0.5\textwidth]{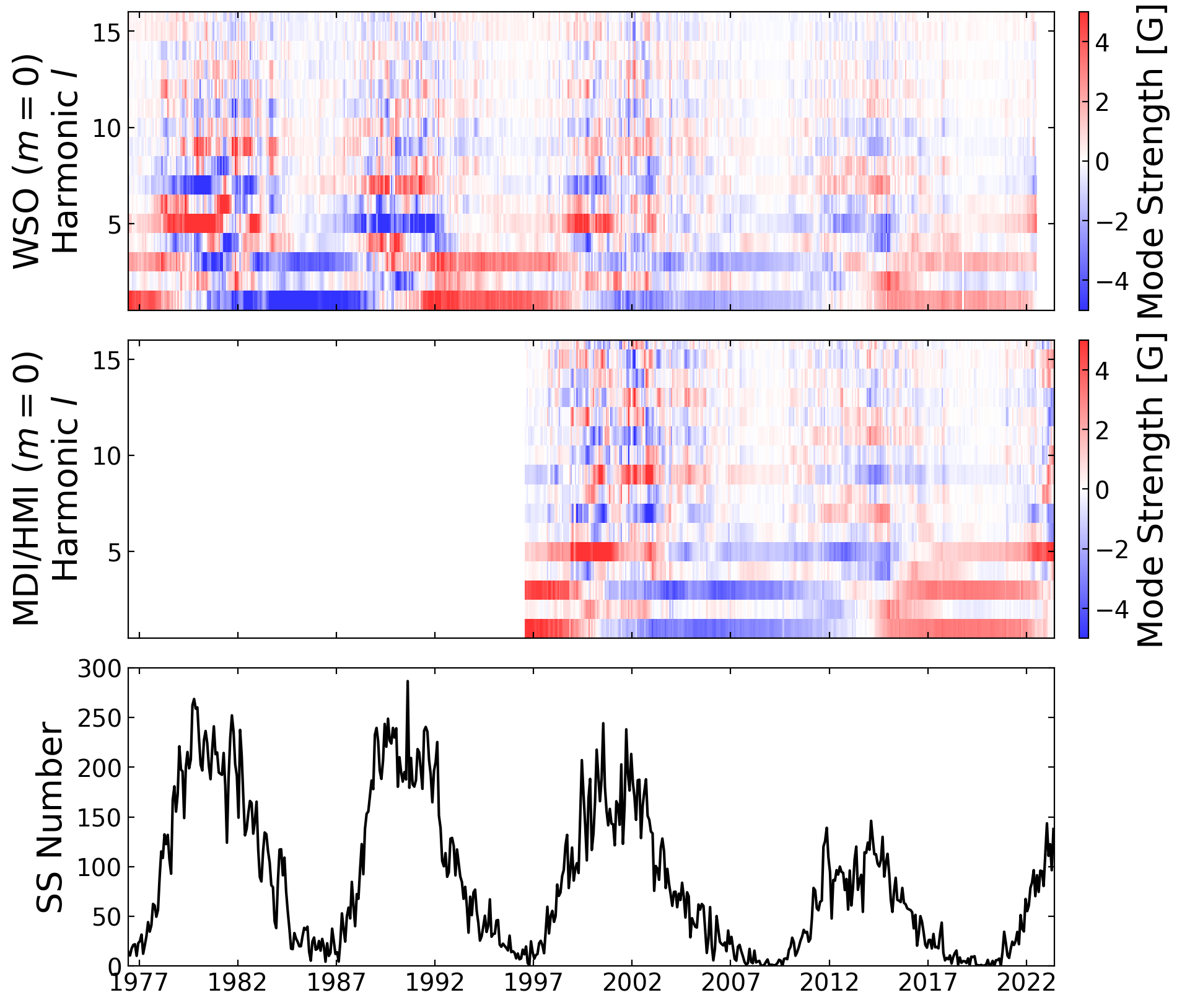}
   \caption{Axial ($m=0$) mode decomposition of the WSO and SOHO/MDI \& SDO/HMI magnetograms, for spherical harmonics up to $l=15$. The top row shows the result from WSO (1976-2022) and second row from SOHO/MDI \& SDO/HMI (1996-present). Field strengths from WSO are multiplied by a factor of 3.5, to be comparable with SOHO/MDI \& SDO/HMI. The bottom row displays the monthly sunspot number during this time, taken from WDC-SILSO.}
   \label{fig:m0modes}
\end{figure}

In order to examine correlations between modes, the axisymmetric $m=0$ field strengths for the lowest 15 modes are plotted in Figure \ref{fig:m0modes}. The top panel contains field strengths from the WSO timeseries, and the second panel contains the SOHO/MDI \& SDO/HMI timeseries. The correlation between modes is most easily identifiable with the odd modes, dipole, octupole, etc. This is due to the sensitivity of the $m=0$ mode to the Sun's polar fields. The correlations progress from low to higher order modes as each cycle progresses.

\begin{figure}
 \centering
  \includegraphics[trim=0cm 0cm 0cm 0cm, clip, width=0.5\textwidth]{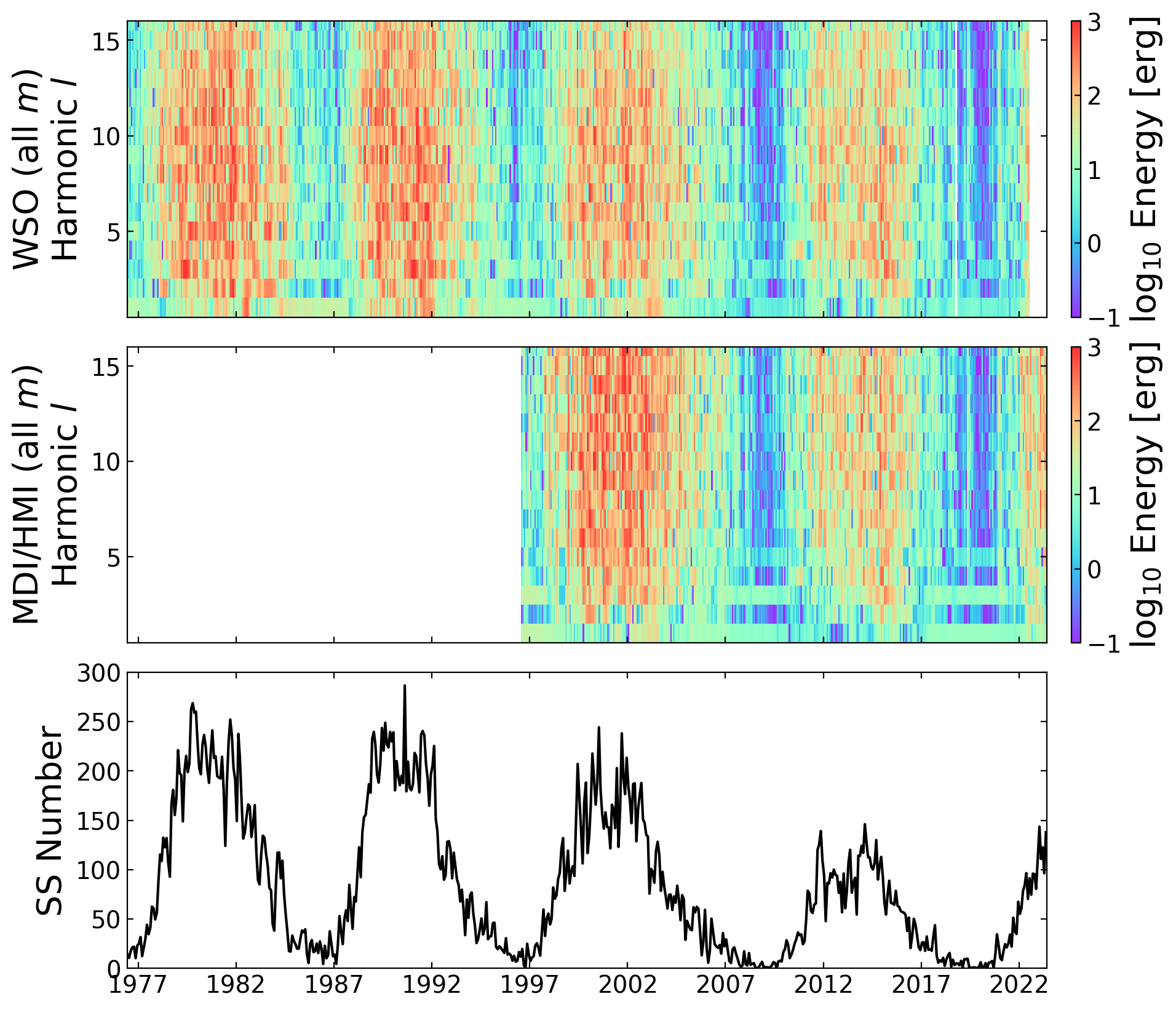}
   \caption{Total energy decomposition (all $m$ components) of the WSO and SOHO/MDI \& SDO/HMI magnetograms, for spherical harmonics up to $l=15$. The top row shows the result from WSO (1976-2022) and second row from SOHO/MDI \& SDO/HMI (1996-present). Field strengths from WSO are multiplied by a factor of 3.5, to be comparable with SOHO/MDI \& SDO/HMI. The bottom row displays the monthly sunspot number during this time, taken from WDC-SILSO.}
   \label{fig:allmodes}
\end{figure}

Figure \ref{fig:allmodes} serves a similar purpose to Figure \ref{fig:m0modes}, now displaying the total energy in each mode (quadratically summing all $m$ components). From this, the cyclic variation of magnetic energy during the solar cycle is visible, with more energy contained in scales associated with sunspots ($l=15-20$), discussed in more detail within \citet{vidotto2016magnetic}. Here the distribution is cut at $l=15$ due to the sensitivity of WSO, which recovers less of the small scale field than its space-based counterparts.

\section{Coloured version of Figure \ref{fig:connect}} \label{ap3}

\begin{figure*}
 \centering
  \includegraphics[trim=0cm 0cm 0cm 0cm, clip, width=\textwidth]{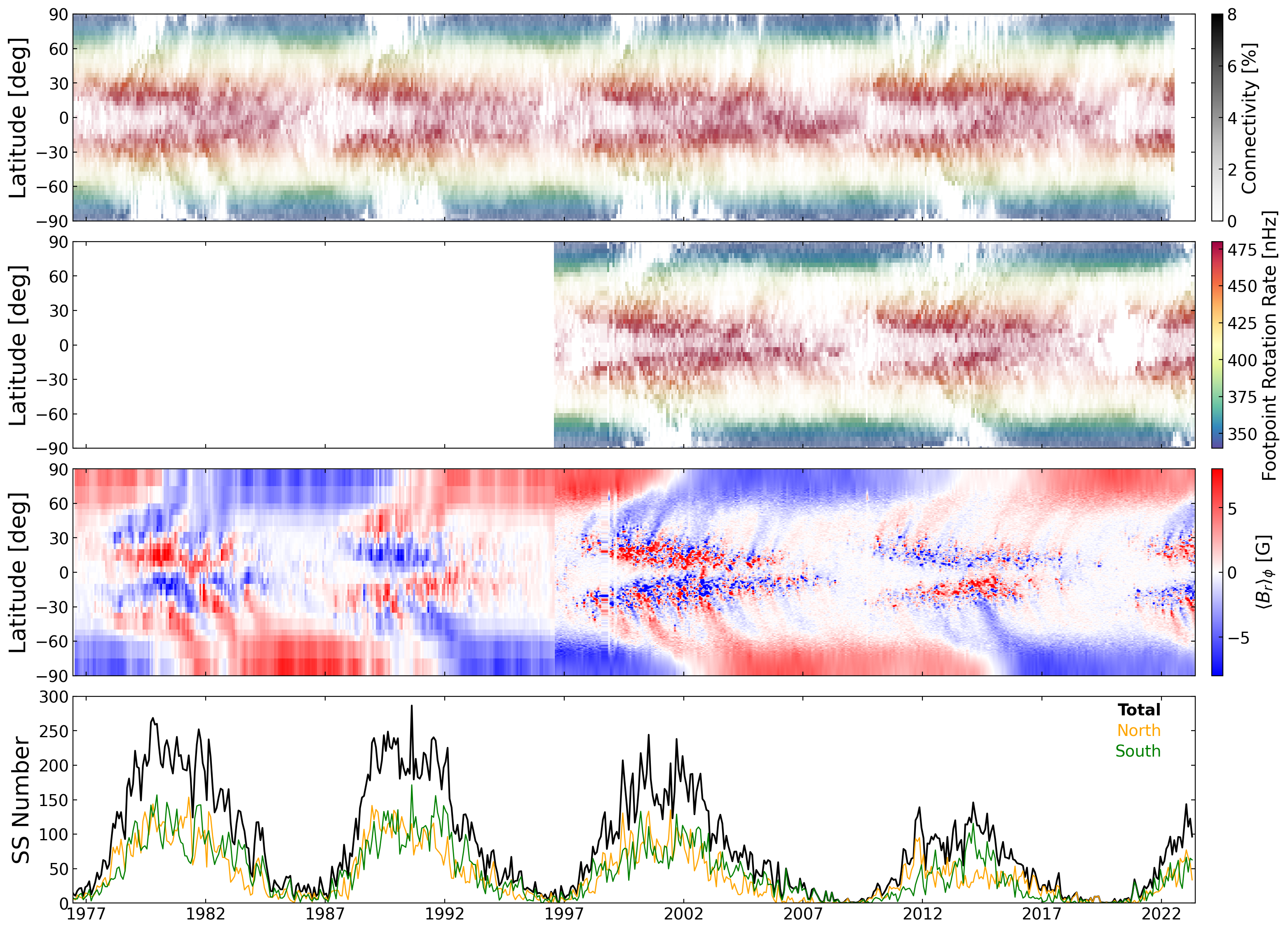}
   \caption{Same as Figure \ref{fig:connect}, but now with the histogram of connectivity coloured by the photospheric rotation rate.}
   \label{fig:connect_color}
\end{figure*}

Figure \ref{fig:connect_color} is similar to Figure \ref{fig:connect}, however with the black and white histogram of PFSS footpoint connectivity replaced by a coloured histogram. The intensity of the colour in the two top panels scales with the connectivity fraction, as in Figure \ref{fig:connect}, but the black histogram colouring is now replaced by colour signifying the surface rotation rate. In this way, the variation between solar wind emerging from the slowly rotating poles, and then the more rapidly rotating active latitudes is more easily visualised. 

\end{appendix}

\end{document}